\documentclass[12pt]{article}

\usepackage{epsfig}

\renewcommand{\arraystretch}{2.0}

\def\dfrac#1#2{{\displaystyle {#1 \over #2}}}
\def\dsum{\mathop{\displaystyle \sum }}

\def\simge{\mathrel{\rlap{\raise 0.511ex \hbox{$>$}}{\lower 0.511ex 
\hbox{$\sim$}}}}
\def\simle{\mathrel{\rlap{\raise 0.511ex \hbox{$<$}}{\lower 0.511ex 
\hbox{$\sim$}}}} 
\def\slash#1{\setbox0=\hbox{$#1$}\dimen0=\wd0                    
      \setbox1=\hbox{/} \dimen1=\wd1 \ifdim\dimen0>\dimen1
      \rlap{\hbox to \dimen0{\hfil/\hfil}} #1                        \else                                       
      \rlap{\hbox to \dimen1{\hfil$#1$\hfil}}
      /   \fi}                                         

\newcommand{\be}{\begin{equation}}
\newcommand{\ee}{\end{equation}}
\newcommand{\bea}{\begin{eqnarray}}
\newcommand{\eea}{\end{eqnarray}}
\newcommand{\nn}{\nonumber}
\newcommand{\la}{\langle}
\newcommand{\ra}{\rangle}
\newcommand{\as}{\alpha_{ s}}

\newcommand{\Heff}{{\cal H}_{ eff}}
\newcommand{\DB}{\Delta B}

\newcommand{\msb}{\mbox{NDR-}\overline{\rm{MS}}}

\newcommand{\ep}{\varepsilon}
\newcommand{\g}{\gamma}

\newcommand{\dgd}{\Delta \Gamma_d/\Gamma_d}
\newcommand{\dgs}{\Delta \Gamma_s/\Gamma_s}
\newcommand{\ds}{\Delta \Gamma_s}
\newcommand{\dd}{\Delta \Gamma_d}

\newcommand{\bdbd}{B_d-\overline B_d}
\newcommand{\bsbs}{B_s-\overline B_s}
\newcommand{\bqbq}{B_q-\overline B_q}
\newcommand{\umcc}{\delta^{cc\,q}_{1/m}}
\newcommand{\umcu}{\delta^{cu\,q}_{1/m}}
\newcommand{\umuu}{\delta^{uu\,q}_{1/m}}
\newcommand{\usmcc}{\delta^{cc\,s}_{1/m}}

\newcommand{\Rdggd}{(2.42\pm 0.59)\times 10^{-3}}
\newcommand{\Rdggs}{(7.4\pm 2.4)\times 10^{-2}}
\newcommand{\Rdgdgs}{(3.2\pm 0.8)\times 10^{-2}}
\newcommand{\Rqpd}{(2.96\pm 0.67)\times 10^{-4}}
\newcommand{\Rqps}{-(1.28\pm 0.27)\times 10^{-5}}

\begin{document}

\begin{titlepage}
\begin{flushright}
RM3-TH/03-9\\
ROME1-1355/03\\
SHEP 0318
\end{flushright}
\vskip 2.4cm
\begin{center}
{\Large \bf Lifetime Differences and CP Violation Parameters \\ \vspace{0.1cm}
of Neutral \boldmath$B$ Mesons \\ \vspace{0.1cm} at the Next-to-Leading Order in QCD \\}
\vskip1.3cm 
{\par\large\bf M.~Ciuchini$^a$, E.~Franco$^b$, V.~Lubicz$^a$,}\\
\vskip0.15cm 
{\par\large\bf F.~Mescia$^c$, and C.~Tarantino$^a$}\\

\vspace{1.cm}
{\normalsize {\sl 
$^a$ Dip. di Fisica, Univ. di Roma Tre and INFN,
Sezione di Roma III, \\
Via della Vasca Navale 84, I-00146 Rome, Italy \\
\vspace{.25cm}
$^b$ Dip. di Fisica, Univ. di Roma ``La Sapienza" and INFN,
Sezione di Roma,\\ P.le A. Moro 2, I-00185 Rome, Italy \\
\vspace{.25cm}
$^c$ Dept. of Physics and Astronomy, Univ. of Southampton, \\
Highfield, Southampton, SO17 1BJ, U.K.
}}
\vspace{.25cm}
\vskip1.5cm
\abstract{ We compute the next-to-leading order QCD corrections to the 
off-diagonal elements of the decay-width matrix $\Gamma$ entering the neutral
$B$-meson oscillations. From this calculation the width differences $\Delta
\Gamma$ and the CP violation parameters $(q/p)$ of $B_d$ and $B_s$ mesons 
are estimated, including the complete ${\cal O}(\as)$ QCD corrections and the 
$1/m_b$ contributions. For the width difference $\ds$ we agree with
previous results. By using the lattice determinations of the relevant hadronic
matrix elements we obtain the theoretical predictions
$\dgd=\Rdggd$ and $\dgs=\Rdggs$. For the CP violation parameters, we find $\vert
(q/p)_d\vert-1=\Rqpd$ and $\vert (q/p)_s\vert-1=\Rqps$. These predictions are 
compatible with the experimental measurements which, however, suffer at present 
from large uncertainties.}

\end{center}
\vspace*{1.cm}
\end{titlepage}

\setcounter{footnote}{0}
\setcounter{equation}{0}

\section{Introduction}
\label{sec:intro}

$B$ physics plays an important role to test and improve our understanding of
the Standard Model (SM). By exploiting the production of $B_d$ mesons with a 
large boost, experiments such as BaBar and Belle can provide accurate
measurements of the decay time distributions and hence valuable information for 
the $B_d$-meson lifetimes, CP violation and mixing parameters. The production of
$B_s$, on the other hand, is out of the reach of the $B$-factories and improved 
measurements of the related physical observables will come from the Run II
at Tevatron and from the LHC. Recently, the measurement of the mass difference
$\Delta m_d$, which controls the
frequency of $B_d$ oscillations, has been further improved. The
world average is now $\Delta m_d=(0.502\pm 0.006)\:\mathrm{ps}^{-1}
$~\cite{Battaglia:2003in}. For $\Delta m_s$ and for the width differences
$\Delta\Gamma_d$ and $\Delta\Gamma_s$, instead, only weak limits exist at
present. Theoretically, the width differences are suppressed by a factor $m^2_b/
m^2_t$ with respect to the corresponding mass differences. In addition, $\Delta
\Gamma_s$ is predicted to be larger than $\Delta \Gamma_d$,
the latter being doubly Cabibbo-suppressed.

The present world average for the width difference of $B_s$ mesons
is~\cite{Battaglia:2003in}
\be
\renewcommand{\arraystretch}{0.6}
\Delta\Gamma_s / \Gamma_s=0.07\begin{array}{l} +0.09 \\ -0.07 \end{array}\,,
\label{eq:dgsmes}
\renewcommand{\arraystretch}{1.0}
\ee
where the constraint $\tau(B_s)=\tau(B_d)$ has been used. Concerning the $B_d$
system, a preliminary result for $\Delta\Gamma_d$ has been recently presented by
the BaBar collaboration~\cite{Aubert:2003se}. The experimental result
reads~\footnote{Note that the definition of $\Delta\Gamma_d$
used in this paper has an opposite sign with respect to one used in
the BaBar paper. In addition, the measurement itself has a sign ambiguity that
can be removed only by measuring the sign of $\cos 2\beta$, where $\beta$ is the phase of the mixing.
This sign is not measured yet but it is known to be
positive in the SM, see e.~g.~\cite{Ciuchini:2003rk}.}
\be
\Delta\Gamma_d / \Gamma_d= 0.008\pm 0.037(\rm{stat.})\pm
0.018(\rm{syst.}).
\label{eq:dgdmes}
\ee

In ref.~\cite{Aubert:2003se}, a preliminary measurement of the parameter
$\vert (q/p)_d\vert$, which defines the amount of CP violation in the mixing, is
also presented
\bea
\label{eq:qpdmes}
\vert (q/p)_d\vert-1=0.029\pm 0.013(\rm{stat.})\pm
0.011(\rm{syst.}).
\eea
In the SM, the CP violation parameters $\vert(q/p)_d\vert-1$ and $\vert (q/p)_s
\vert-1$ are suppressed, with respect to the width differences, by a factor 
$m_c^2/m_b^2$. Moreover, $|(q/p)_s|-1$ has an additional suppression factor
$\lambda^2$ (where $\lambda$ is the sine of the Cabibbo angle) with respect to $|(q/p)_d|-1$.

The phenomenology of $\bqbq$ oscillations is described in terms of a $2\times 2$
effective Hamiltonian matrix, $M^q-i\Gamma^q/2$. The b-quark mass, being large compared
to $\Lambda_{QCD}$, allows to apply an operator product expansion (OPE) to
the calculation of the decay-width matrix
$\Gamma^q$ ~\cite{Khoze:1987fa,ope}. In the
case of the off-diagonal matrix elements, which the observables $\Delta\Gamma$
and $(q/p)$ are related to, the leading term in this expansion is represented by
the so-called ``spectator effect'' contributions.

Recently, spectator effects have been computed at ${\cal O}(\as)$ and ${\cal O}
(\Lambda_{QCD}/m_b)$ for the B-meson and $\Lambda_b$ 
lifetimes~\cite{Ciuchini:2001vx}-\cite{Gabbiani:2003pq} and for the Cabibbo-favoured
transition ($\bar b s \to c \bar c$) contributing to $\Delta
\Gamma_s$~\cite{Beneke:1996gn,Beneke:1998sy}. In the case of $\Delta\Gamma_d$, 
the complete set of the $\Lambda_{QCD}/m_b$ contributions has been presented 
in~\cite{Dighe:2001gc}, but the QCD next-to-leading order (NLO) corrections have
been only estimated in that paper in the limit of vanishing charm quark mass, 
using the results of~\cite{Beneke:1998sy}.

The main result of this paper is the calculation of NLO QCD corrections to 
$\Gamma^d_{21}$ including the contributions from a non-vanishing charm quark 
mass. Our results for the NLO Wilson coefficients of $\Gamma^s_{21}$ 
agree with ref.~\cite{Beneke:1998sy} and in addition we have computed the
Cabibbo-suppressed contributions. We have also checked the computation of the
$\Lambda_{QCD}/m_b$ corrections to $\Gamma^s_{21}$ and $\Gamma^d_{21}$ at the
LO in QCD and the results agree with refs.~\cite{Beneke:1996gn,Dighe:2001gc}.
Finally, we have computed the matching between the QCD and HQET $\DB=2$
operators at the NLO in QCD and found agreement with the results presented in
ref.~\cite{Becirevic:2001xt}. As phenomenological applications of our
calculations, by using the lattice determinations of the relevant hadronic 
matrix elements~\cite{Becirevic:2001xt}, we estimate the width differences
$\Delta\Gamma_d$, $\Delta\Gamma_s$ and the CP violation parameters $(q/p)_d$ and
$(q/p)_s$.

The inclusion of NLO corrections is important in order to match the scale and 
scheme dependence of the renormalized operators entering the OPE with that of
the Wilson coefficients, thus reducing the theoretical uncertainties.
In addition, the NLO
corrections are found to be generally large. Taking
into account the finite value of the charm quark mass is also an important issue
for a proper NLO estimate of $\Delta\Gamma_d$ and of the CP violation parameters
$\vert(q/p)_d\vert$ and $\vert(q/p)_s\vert$. In the limit $m_c\to 0$, $\Delta
\Gamma_d$ becomes independent of the phase of the mixing amplitude ($2\beta$ in
the SM) and CP is not violated in the mixing, {\it i.e.}~$\vert(q/p)_d\vert$ and
$\vert(q/p)_s\vert$ are simply equal to one~\cite{hagelin}. Therefore, the
dependence on the mixing phase as well as the CP violating terms disappear from
the NLO corrections in the vanishing charm mass limit.

Our theoretical estimates of the width differences, obtained by using the
lattice determinations of the relevant hadronic matrix
elements~\cite{Becirevic:2001xt}, are
\be
\label{eq:nlores}
\dgs=\Rdggs\,,\quad\dgd=\Rdggd\,.
\ee
These results can only be compared at present with the experimental
determinations given in eqs.~(\ref{eq:dgsmes}) and~(\ref{eq:dgdmes}).  The
predictions are consistent with the measured values but, given the large
experimental uncertainties, it is certainly premature to draw any conclusion. 
Our estimates are accurate at the NLO in $\alpha_s$ for the leading term in the
$1/m_b$ expansion and at the LO for the first power-correction.

It is interesting to consider also the estimate of the ratio $\dd/\ds$, for
which we obtain the prediction
\be
\label{eq:nlores5}
\dd/\ds= \Rdgdgs\,.
\ee 
In this ratio the uncertainties coming from higher orders of QCD and $\Lambda_
{QCD}/m_b$ corrections, as well as from the non perturbative estimates of the 
hadronic matrix elements, cancel out to some extent.

Finally, our predictions for the CP violation parameters are
\be
\label{eq:nlores4}
\left\vert\left(q/p\right)_d\right\vert-1=\Rqpd\,,\quad
\left\vert\left(q/p\right)_s\right\vert-1=\Rqps\,.
\ee
The value of $\left\vert\left(q/p\right)_d\right\vert$ can be compared with the
preliminary experimental determination given in eq.~(\ref{eq:qpdmes}). Improved
measurements are certainly needed to make this comparison more significant.

We conclude this section by presenting the plan of this paper. In
sect.~\ref{sec:formulae} we review the basic formalism of the $\bqbq$ mixing,
introducing all the physical quantities we are interested in. The details of the
NLO calculation are presented in sect.~\ref{sec:results}, along with the
calculation of  the  $\Lambda_{QCD}/m_b$ correction. In sect.~\ref{sec:phenom}
we present the theoretical predictions for the observables under consideration
in both the $\bdbd$ and $\bsbs$ systems. Finally, the analytical expressions for
the matching coefficient functions are given in appendix A, while detailed
definitions of the renormalization schemes used in the calculation and results
for the matching
between QCD and HQET operators can be found in appendix B.

\section{\boldmath$\bqbq$ Basic Formalism}
\label{sec:formulae}

The neutral $B_d$ and $B_s$ mesons mix with their antiparticles leading to 
oscillations between the mass eigenstates. The time evolution of the neutral 
mesons doublet is described by a Schroedinger equation with an effective
$2 \times 2$ Hamiltonian
\be
i\frac{d}{dt} \pmatrix{B_q\cr {\overline B}_q}=
\left[ \pmatrix{M_{11}^q & {M_{21}^q}^*\cr M_{21}^{q} & M_{11}^q}
-\frac{i}{2} \pmatrix{\Gamma_{11}^q & {\Gamma_{21}^q}^*\cr \Gamma_{21}^{q}& 
\Gamma_{11}^q} \right]\pmatrix{B_q\cr {\overline B}_q}\,.
\label{eq:schro}
\ee
The mass difference $\Delta m_q$ and the width difference $\Delta \Gamma_q$ are 
defined as
\be
\Delta m_q=m^q_H-m^q_L\,,\quad
\Delta\Gamma_q=\Gamma^q_L-\Gamma^q_H\,,
\ee
where $H$ and $L$ denote the Hamiltonian eigenstates with the heaviest and
lightest mass eigenvalue respectively. These states can be written as
\be
\vert B_q^{L,H}\rangle ={1\over \sqrt{1+\vert (q/p)_q \vert^2}}\,\left(
\vert B_q\rangle \pm  \left(q/p\right)_q\vert {\overline B}_q\rangle
\right)\,.
\ee
The phase of $(q/p)_q$ depends on the phase convention of the $B$ states and
hence it is not measurable by itself. In this paper, we are interested in
$\vert (q/p)_q\vert$ only.

Theoretically, the experimental observables $\Delta m_q$, $\Delta \Gamma_q$ and
$\vert(q/p)_q\vert$ are related to $M^q_{21}$ and $\Gamma^q_{21}$ in
eq.~(\ref{eq:schro}) as follows~\footnote{For more information about the basic
definitions in the $\bdbd$- and $\bsbs$-mixing, see for
instance~\cite{Battaglia:2003in} and references therein.}
\be
\begin{array}{c}
(\Delta m_q)^2 -\dfrac{1}{4}(\Delta \Gamma_q)^2=4\vert M^q_{21}\vert^2 -
\vert \Gamma^q_{21}\vert^2,\nn\cr
\Delta m_q\, \Delta \Gamma_q=-4\, \mathrm{Re}\left(M^{q\,*}_{21}\,\Gamma^q_{21}
\right)\,, \quad\,
\left\vert\left({q/p}\right)_q\right\vert=\left\vert\sqrt{
\dfrac{2 M^q_{21}-i\Gamma^q_{21}}{2 M^{q\,*}_{21}-i\Gamma^{q\,*}_{21}}}
\right\vert\,.
\label{eq:dgamma}
\end{array}
\ee
In the $\bdbd$ and $\bsbs$ systems, the ratio $\Gamma^q_{21}/M^q_{21}$ is of
${\cal O}(m_b^2/m_t^2)\simeq 10^{-3}$. Therefore, by neglecting terms of ${\cal 
O}(m_b^4/m_t^4)$, eq.~(\ref{eq:dgamma}) can be simply written as
\be
\Delta m_q= 2\,\vert M^q_{21}\vert\,,\quad
\Delta \Gamma_q=-2\,\vert M^q_{21}\vert\,
{\mathrm{Re}}\left(\dfrac{\Gamma^q_{21}}{M^{q}_{21}}\right)\,,\quad
\left\vert\left(q/p\right)_q\right\vert=1+\dfrac{1}{2}\,{\mathrm{Im}}\left(
\dfrac{\Gamma^q_{21}}{M^q_{21}}\right)\,.
\label{eq:dgammared}
\ee

The matrix elements $M^q_{21}$ and $\Gamma^q_{21}$ are related to the dispersive
and the absorptive parts of the $\DB=2$ transitions respectively. In the SM,
these transitions are the result of second-order charged weak interactions 
involving the well-known box diagrams.

The quantity $M^q_{21}$ has been computed, at the NLO in QCD, in
ref.~\cite{Buras:1990fn} and it is given by
\bea
M^q_{21} = \frac{G_F^2 M_W^2 \eta_B }{(4 \pi)^2 (2 M_{B_q} )
      }(V_{tb}^* V_{tq})^2
       S_0(x_t)\, \langle \overline B_q \vert
(\bar b_i q_i)_{V-A} (\bar b_j q_j)_{V-A})
\vert B_q \rangle\,,
\label{eq:mq21}
\eea
where $x_t = \bar{m}_t^2/M_W^2 $, $\eta_B $ is the QCD correction factor and 
$S_0$ is the Inami--Lim function. Here and in the following we use the notation 
$(\bar q q)_{V\pm A}=\bar q \gamma_\mu(1\pm\gamma_5) q$ and $(\bar q q)_{S\pm P}
=\bar q (1\pm\gamma_5) q$. A sum over repeated colour indices is always 
understood.

The matrix element $\Gamma^q_{21}$  can be
written as
\be
\Gamma^q_{21} =
\frac{1}{2 M_{B_q}} \mathrm{Disc} \langle \overline B_q \vert  i \int d^4x \;
T \left( \Heff^{\DB=1}(x) \Heff^{\DB=1}(0) \right) \vert B_q \rangle\, ,
\label{eq:master}
\ee
where ``$\mathrm{Disc}$'' picks up the discontinuities across the physical cut
in the time-ordered product of the effective Hamiltonians.

The $\DB=1$ effective Hamiltonian relevant to $b\to d$ and $b\to s$ transitions
is
\bea
\label{eq:hdb1}
&&\Heff^{\DB=1} =
\frac{G_F}{\sqrt{2}} \Bigg\{ \Bigg[\Bigg( V^\ast_{cb} V_{ud} \left( C_1 Q_1 + 
C_2 Q_2 \right) + V^\ast_{cb} V_{cd} \left( C_1 Q^c_1 + C_2 Q^c_2 \right) + 
\left( c \leftrightarrow u \right) \Bigg) -  \nn \\
&&\qquad V^\ast_{tb} V_{td} \left( \sum_{i=3}^{6} C_i Q_i + C_{8G} Q_{8G} 
\right) \Bigg] + \Bigg[ d \to s \Bigg] \Bigg\} + h.c.\,.
\label{eq:heffdb1}
\eea 
The $C_i$ are the $\DB=1$ Wilson coefficients, known at the NLO in perturbation
theory~\cite{nlodb1a}-\cite{nlodb1c}, and the operators $Q_i$ are defined as
\be
\begin{array}{ll}
Q_1= (\bar b_i c_j)_{V-A} (\bar u_j d_i)_{V-A}\,, \qquad \qquad
& Q_2= (\bar b_i c_i)_{V-A} (\bar u_j d_j)_{V-A}\,, \\
Q^c_1= (\bar b_i c_j)_{V-A} (\bar c_j d_i)_{V-A}\,, \qquad \qquad
& Q^c_2= (\bar b_i c_i)_{V-A} (\bar c_j d_j)_{V-A}\,, \\
Q_3= (\bar b_i d_i)_{V-A} \dsum\limits_{q} (\bar q_j q_j)_{V-A}\,, \qquad 
\qquad
& Q_4= (\bar b_i d_j)_{V-A} \dsum\limits_{q} (\bar q_j q_i)_{V-A}\,, \\
Q_5= (\bar b_i d_i)_{V-A} \dsum\limits_{q} (\bar q_j q_j)_{V+A}\,, \qquad 
\qquad
& Q_6= (\bar b_i d_j)_{V-A} \dsum\limits_{q} (\bar q_j q_i)_{V+A}\,, \\
Q_{8G}= \dfrac{g_s}{8\pi^2} m_b \bar b_i \sigma^{\mu\nu} 
\left(1-\gamma^5\right) t^{a}_{ij} d_j G^{a}_{\mu\nu}\,.
\end{array}
\label{eq:operatori}
\ee
Due to the large mass of the $b$ quark, the time-ordered product in 
eq.~(\ref{eq:master}) can be expanded in a sum of local operators of increasing 
dimension~\cite{Khoze:1987fa,ope}. Up to order $1/m_b$, this expansion reads
\be
\Gamma^q_{21} =
-\dfrac{G_F^2 m_b^2}{24 \pi M_{B_q}}\left[
c^q_1(\mu_2) {\langle \overline B_q \vert {\cal O}^q_1(\mu_2) \vert B_q
\rangle}
+c^q_2(\mu_2) {\langle \overline B_q \vert {\cal O}^q_2(\mu_2) \vert B_q
\rangle} + \delta^q_{1/m}\right]\, ,
\label{eq:gamma12q}
\ee
where 
\be
{\cal O}^q_1= (\bar b_i q_i)_{V-A} (\bar b_j q_j)_{V-A}\,,\qquad 
{\cal O}^q_2= (\bar b_i q_i)_{S-P} (\bar b_j q_j)_{S-P}\,,
\label{eq:effop}
\ee
and $\delta^q_{1/m}$ denotes the sub-leading $1/m_b$ corrections. Thus, the OPE
of $\Gamma^q_{21}$ at the leading order is expressed in terms of only two local
four-fermion operators, ${\cal O}^q_1$ and ${\cal O}^q_2$.

\begin{figure}[t]
\begin{center}
\includegraphics[width=6cm]{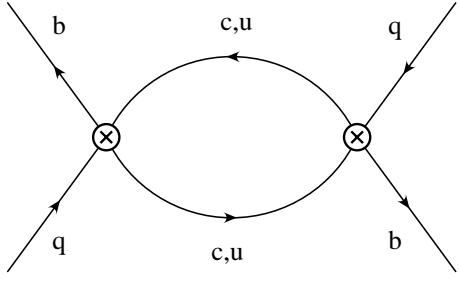}
\end{center}
\caption{\it Feynman diagram contributing to $\Gamma_{21}^q$ at the LO in
QCD.}
\label{fig:unosum}
\end{figure}
In terms of the CKM matrix elements $V^*_{cb}V_{cq}$ and $V^*_{tb}V_{tq}$ (the latter
regulates the $M^q_{21}$ contribution), the Wilson coefficients $c^q_i$ and the
$\delta^q_{1/m_b}$ in eq.~(\ref{eq:gamma12q}) can be written as
\be
\begin{array}{l}
c^q_i=
(V^*_{tb}V_{tq})^2\,D^{uu}_i
+ 2\,V^*_{cb}V_{cq}V^*_{tb}V_{tq}\,\left( D^{uu}_i - D^{cu}_i \right)
+(V^*_{cb}V_{cq})^2\,\left( D^{cc}_i + D^{uu}_i -2 \,D^{cu}_i\right) \,, \cr
\delta^q_{1/m}=
(V^*_{tb}V_{tq})^2\,\umuu
+ 2\,V^*_{cb}V_{cq}V^*_{tb}V_{tq}\,\left( \umuu - \umcu \right)
+(V^*_{cb}V_{cq})^2\,\left( \umcc + \umuu -2 \,\umcu\right)
\end{array}
\label{eq:dccs}
\ee
where the labels $cc,\,cu$ and $uu$ denote the intermediate quark pair
contributing to $\Gamma_{21}^q$, see the diagram in fig.~\ref{fig:unosum}. The
unitarity of the CKM matrix has been used to drop out $V^*_{ub}V_{uq}=-
V^*_{cb}V_{cq} - V^*_{tb}V_{tq}$.

In this paper, we compute the QCD NLO corrections to $D^{cc}_i\,,D^{cu}_i\,,
D^{uu}_i$ taking into account the finite value of the charm quark mass. Our 
results are collected in appendix A. The results for $D^{cc}_i$ are in agreement
with a previous calculation~\cite{Beneke:1998sy}. The functions $D^{cu}_i\,,
D^{uu}_i$ have been previously estimated~\cite{Dighe:2001gc} taking the limit of
a massless charm quark of $D^{cc}_i$. As stressed in the introduction, this 
approximation is not fully satisfactory, in particular if one wants to 
investigate the dependence on the phase of the mixing  in $\dgd$ and
the $|q/p|$ parameters.

As far as the $1/m$ contributions are concerned, $\umcc$ and $\umcu$, $\umuu$
in eq.~(\ref{eq:dccs}) have been calculated
in~\cite{Beneke:1996gn} and~\cite{Dighe:2001gc} respectively.
We have repeated the calculation by expanding
in $1/m_b$ the relevant Feynman diagrams and found agreement with the
previous results.

\section{Calculation of the  Wilson coefficients for \boldmath$\Gamma^q_{21}$}
\label{sec:results}
In this section, we discuss the basic ingredients of the calculation of the NLO
QCD corrections to $\Gamma^q_{21}$, at the leading power in $1/m_b$, and of the
calculation of the $1/m_b$ corrections at the LO in QCD. We refer the interested
reader to refs.~\cite{Ciuchini:2001vx,Franco:2002fc} for further details on the 
NLO calculation.

\subsection{NLO QCD corrections at the leading order in \boldmath$1/m_b$}
Although at the leading order in $1/m_b$ the heavy quark expansion of the width
difference could be expressed directly in terms of QCD operators, we found it easier to perform
first the expansion in terms of
HQET operators and then translating the results into QCD, once IR divergences
are canceled out.

\begin{figure}[t]
\begin{center}
\includegraphics[width=\textwidth]{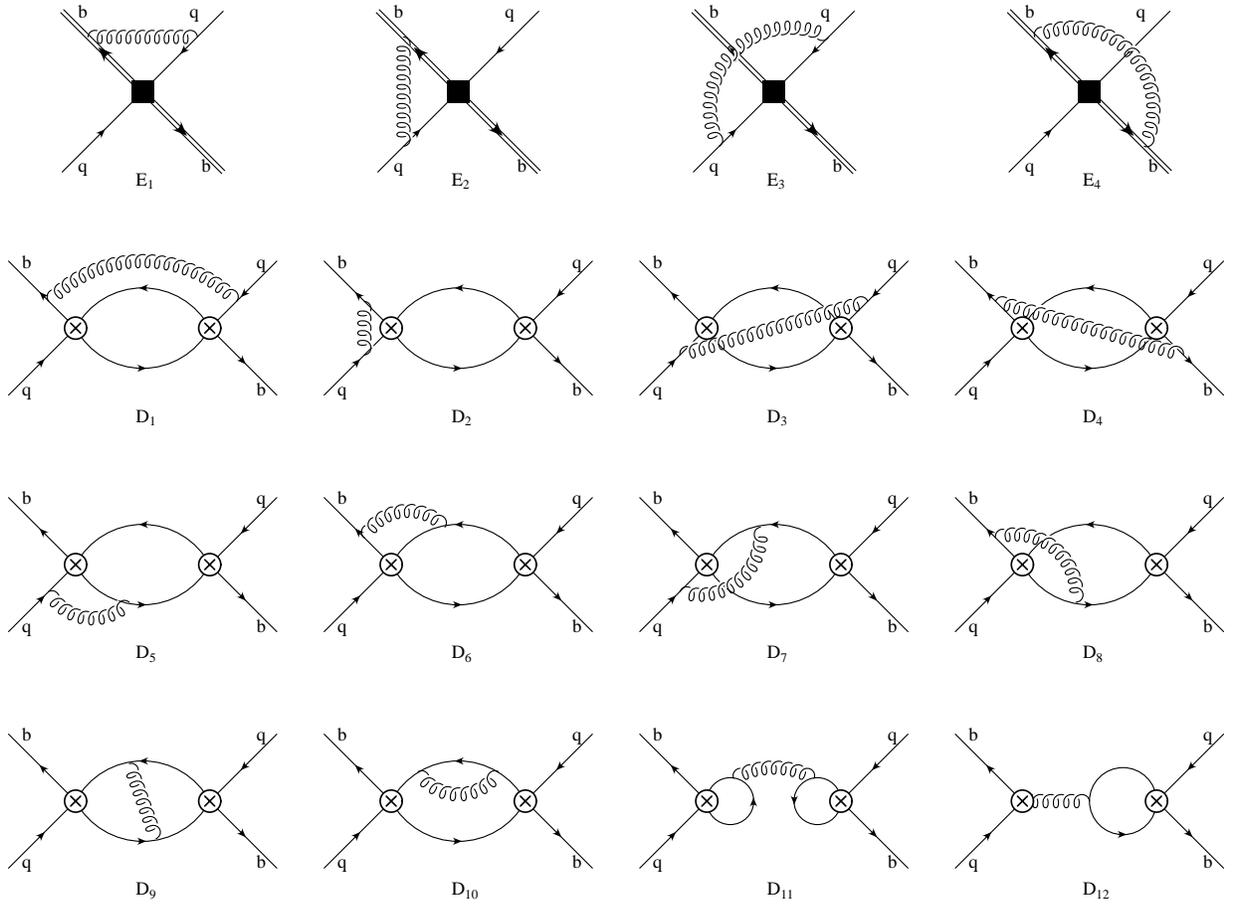}
\end{center}
\caption{\it Feynman diagrams which contribute at the NLO to the matrix element
of the transition operator ${\cal T}$, both in the full theory ($D_i$) and in
the effective theory ($E_i$). The diagrams obtained from $E_1,\,E_2,
\,D_1,\,D_2,\,D_5,\,D_6,\,D_7,\,D_8,\,D_{10}$ and $D_{12}$ by a $180^o$
rotation (keeping fixed the flavour of the internal lines), as well as those
containing the self-energy one-loop corrections to
the external fields are not shown.}
\label{fig:fddb1}
\end{figure}
In order to compute the Wilson coefficients of the $\DB=2$ operators at the NLO,
we have evaluated in QCD the imaginary part of the diagrams $D_i$ shown in
fig.~\ref{fig:fddb1} ({\it full} theory) and in the HQET the diagrams $E_i$
({\it effective} theory). The external quark states have been taken on-shell and
all quark masses, except $m_b$ and $m_c$, have been neglected. More 
specifically, we have chosen the heavy quark momenta $p_b^2=m_b^2$ in QCD and 
$k_b=0$ in the HQET while, for the external light quarks, $p_q=0$ in both cases.
We have performed the calculation in a generic covariant gauge, in order to 
check the gauge independence of the final results. Two-loop integrals have been
reduced to a set of independent master integrals by using the TARCER 
package~\cite{tarcer}.

IR and UV divergences have been regularized by $D$-dimensional regularization 
with anticommuting $\gamma_5$ (NDR). As discussed in detail in 
ref.~\cite{Ciuchini:2001vx}, in the presence of dimensionally-regularized IR 
divergences the matching must be consistently performed in $D$ dimensions, 
including the contribution of (renormalized) evanescent operators~\cite{Misiak:1999yg}. All
the evanescent operators entering the matching procedure are collected in
appendix B.

As a check of the perturbative calculation, we have verified that our results
for the Wilson coefficients satisfy the following requirements:
\begin{itemize}
\item {\it gauge invariance}:
the coefficient functions in the $\overline{\rm{MS}}$ scheme are explicitly
gauge-invariant. The same is true for the full and the effective amplitudes
separately;
\item {\it renormalization-scale dependence}:
the coefficient functions have the correct logarithmic scale dependence as
predicted by the LO anomalous dimensions of the $\DB=2$ and $\DB=1$ operators;
\item {\it renormalization-scheme dependence}:
we performed the calculation in two different $\msb$ schemes for the $\DB=2$
operators (see appendix B and refs.~\cite{Beneke:1998sy,db2nlo1}
for a detailed definition of
these schemes) and verified that the NLO Wilson coefficients obtained in the
two cases are related by the appropriate matching matrix;
\item {\it IR divergences}:
the coefficient functions are infrared finite;
\item{\it IR regularization}:
in the limit $m_c\to 0$, we have also checked that a calculation using the gluon
mass as IR regulator gives the same results. In this case, the matrix elements
of the renormalized evanescent operators vanish and hence these operators do not
give contribution in the matching procedure;
\item{{\it comparison with the results in}~\cite{Beneke:1998sy}}:
our results for $D^{cc}_i$ defined in eq.~(\ref{eq:dccs}) and given in
appendix~A agree with those obtained in ref.~\cite{Beneke:1998sy}.
\end{itemize}

The $\DB=2$ effective theory is derived from the double insertion of the $\DB=1$
effective Hamiltonian. Therefore, the coefficient functions $D^{cc}_k$, $D^{cu}
_k$ and $D^{uu}_k$ of the $\DB=2$ effective theory depend quadratically on the
coefficient functions $C_i$ of the $\DB=1$ effective Hamiltonian. The dependence
on the renormalization scheme and on the scale $\mu_1$ of the $\DB=1$ operators
actually cancels order by order in perturbation theory against the corresponding
dependence of the $\DB=1$ Wilson coefficients $C_i$. Therefore, the coefficient
functions $c^q_k$ only depend on the renormalization scheme and on the scale $\mu_2$
of the $\DB=2$ operators.

\begin{table} [t]
\begin{center}
\begin{tabular}{||l||c||c|c||c|c||}
\cline{2-6}
\multicolumn{1}{c||}{}       & LO    & ${\cal O}(\as)$&NLO&${\cal
O}(\as)$&NLO\\
\multicolumn{1}{c||}{}       &       &\multicolumn{2}{c||}{$\as m_c=0$}& & \\
\hline
$D^{uu}_1$                          & 0.260 & -0.123 & 0.174 & -0.123 & 0.174\\
$D^{uu}_1 - D^{cu}_1$               & 0.103 & 0 & 0.105& -0.034  & 0.072 \\
$D^{cc}_1 + D^{uu}_1 -2 \,D^{cu}_1$ & 0.004  & 0 & 0.004& -0.005 & -0.001\\
\hline
$D^{uu}_2$                          & -1.56 & 0.44 & -1.00& 0.44 &-1.00  \\
$D^{uu}_2 - D^{cu}_2$               & -0.035 & 0 & -0.033& 0.012  &-0.022 \\
$D^{cc}_2 + D^{uu}_2 -2 \,D^{cu}_2$ & 0.016  & 0 & 0.015&-0.014 & 0.001 \\
\hline
\end{tabular}
\end{center}
\caption{\it Values of the Wilson coefficients defined in eq.~(\ref{eq:dccs})
computed at the reference scales $\mu_1=\mu_2=m_b=4.75$ GeV. The LO (first
column) and NLO (third and fifth column) coefficients are shown together with
the pure ${\cal O}(\as)$ contribution (second and fourth column). The results
given in the second and third column are those obtained by neglecting the
${\cal O}(\as m_c)$ corrections.}
\label{tab:coeffQCD}
\end{table}
The analytical expressions of the Wilson coefficients are collected in
appendix~A. For illustrative purposes, the corresponding numerical values,
at the reference scales
$\mu_1=\mu_2=m_b$ are presented in table~\ref{tab:coeffQCD}. In this table,
the LO and NLO coefficients are shown together with the pure
${\cal O}(\as)$ contribution, both in the case of a finite charm quark mass and
in the charm massless limit~\footnote{Note that, in what we call the pure
${\cal O}(\as)$ contribution in table~\ref{tab:coeffQCD}, the NLO corrections to
the $\Delta B=1$ Wilson coefficients are not included. For this reason, the
NLO results in the table~\ref{tab:coeffQCD} differ from the sums of the
LO coefficients and the pure ${\cal O}(\as)$ contributions.}. The NLO
coefficients refer to the QCD operators renormalized in the $\msb$ scheme of
ref.~\cite{Beneke:1998sy}. In order to obtain the numerical results we used the
central values of the input parameters given in table~\ref{tab:inputs}.

By looking at the results shown in table~\ref{tab:coeffQCD}, we see that the NLO
corrections are large and the effect of charm contributions is crucial to
estimate accurately all the terms in eq.~(\ref{eq:dccs}).

\subsection{\boldmath$1/m_b$ Corrections to \boldmath$\Gamma_{21}^q$ at the
LO in QCD}
In order to compute the $1/m_b$ correction at the LO in QCD, the imaginary part of the
diagrams in fig.~\ref{fig:unosum} have been evaluated between on-shell quark
states and expanded in the quark momenta~\footnote{In the heavy quark expansion
we consider the strange quark mass of ${\cal O}(\Lambda_{QCD})$, whereas $m_u$
and $m_d$ are neglected.}.
The result of the
{\em full theory} has been then matched at ${\cal O}(1/m_b)$ onto the following
independent set of QCD operators,
\be
\begin{array}{ll}
R^q_1 =\dfrac{m_q}{m_b}(\bar b_i q_i)_{S-P}(\bar b_j q_j)_{S+P}\,,
& \! \! \! \!\! \! \!
R^q_2 =\dfrac{1}{m^2_b}(\bar b_i {\overleftarrow D}_{\!\rho}
\gamma^\mu(1-\gamma_5)D^\rho q_i)( \bar b_j\gamma_\mu(1-\gamma_5)q_j)\,, \cr
R^q_3 =\dfrac{1}{m^2_b}(\bar b_i{\overleftarrow D}_{\!\rho}
(1-\gamma_5)D^\rho q_i) (\bar b_j(1-\gamma_5)q_j)\,, &
R^q_4 =\dfrac{1}{m_b}(\bar b_i(1-\gamma_5)iD_\mu q_i)
(\bar b_j\gamma^\mu(1-\gamma_5)q_j)\,.
\label{rrt1}
\end{array}
\ee
Since derivatives acting on $b$ fields scale as $m_b$, the above operators are
manifestly of order $1/m_b$.
The results for the $1/m_b$ terms defined in eq.~(\ref{eq:dccs}) read
\bea
\umcc&=& \sqrt{1 - 4\,z}\,( (1 + 2 z)\left[K_2\, (\la R^q_2\ra + 2\,\la R^q_4
\ra) -2\, K_1\, (\la R^q_1\ra+\la R^q_2\ra)\right]\nn\\
& -&
\dfrac{12\, z^2}{1 - 4\,z}\left[K_1\,(\la R^q_2\ra+2\,\la R^q_3\ra) + 2\,K_2\,
\la R^q_3\ra\right])\,,\nn\\
\umcu&=& \left(1 - z\right)^2((1 + 2 z)\left[K_2\, (\la R^q_2\ra + 2\,\la
R^q_4\ra) -2\, K_1\, (\la R^q_1\ra+\la R^q_2\ra)\right]\\
& -&
\dfrac{6\, z^2}{(1 - z)}(K_1\,(\la R^q_2\ra+2\,\la R^q_3\ra) + 2\,K_2\, \la
R^q_3\ra))\,,\nn\\
\umuu&=&\left[K_2\, (\la R^q_2\ra + 2\,\la R^q_4\ra) -2\, K_1\, (\la R^q_1\ra+
\la R^q_2\ra)\right]\,,\nn
\eea
where $K_1$ and $K_2$ are related to the $\Delta B=1$ coefficients,
$K_1=3\, C^2_1 + 2\, C_1\,C_2$ and $K_2=C^2_2$, and $z=(m_c/m_b)^2$.
Our results agree with previous
calculations~\cite{Beneke:1996gn,Dighe:2001gc}~\footnote{
For the reader's benefit, we observe that the expression of $\umcc$ in eq.~(49)
of ref.~\cite{Dighe:2001gc} has been expanded up to ${\cal O}(z^3)$.
}.

\section{Width differences and CP violation parameters in \boldmath$\bqbq$
mixing}
\label{sec:phenom}

In this section we present the theoretical estimates for the quantities of
interest in this paper: the width differences $\dgd$ and $\dgs$ and the CP
violation parameters $\vert (q/p)_d\vert$ and $\vert (q/p)_s\vert$.

Our estimates neglect terms of ${\cal O}(\alpha_s^2)$, ${\cal O}(\alpha_s\,
(\Lambda_{QCD}/m_b))$ and ${\cal O}((\Lambda_{QCD}/m_b)^2)$, whereas the charm
quark mass contributions have been fully taken into account. For the relevant
hadronic matrix elements we have used the lattice determination of
ref.~\cite{Becirevic:2001xt}.

In eq.~(\ref{eq:dgammared}), the width difference $\Delta\Gamma_q$ and the
parameter $\vert (q/p)_q\vert$ for neutral $B$ mesons are expressed in terms
of the real and the imaginary part of $\Gamma^q_{21}/M^q_{21}$ respectively.
Taking into account the expressions obtained in eq.~(\ref{eq:mq21}) for $M^q_
{21}$ and eqs.~(\ref{eq:gamma12q})-(\ref{eq:dccs}) for $\Gamma^q_{21}$, we can
write

\bea
\label{eq:dgm}
\Delta \Gamma_q &=& \dfrac{\Delta m_q \,{\cal K}}{\langle \overline B_q \vert
{\cal O}^q_1 \vert B_q\rangle}\times\nn\\
&&\left(\dsum\limits_{i=1,2}\langle \overline B_q \vert {\cal O}^q_i \vert B
_q\rangle \left[ D^{uu}_i +\dfrac{\cos 2\beta_q}{R^2_{t\,q}}\,\left( D^{cc}_i
+ D^{uu}_i -2 \,D^{cu}_i\right) - 2\,\dfrac{\cos\beta_q}{R_{t\,q}}\,\left(
D^{uu}_i - D^{cu}_i \right)\right]\right.\nn\\
&&+\left.
\left[ \umuu +\dfrac{\cos 2\beta_q}{R^2_{t\,q}}\,\left(\umcc
+ \umuu -2 \,\umcu\right)
- 2\,\dfrac{\cos\beta_q}{R_{t\,q}}\,\left( \umuu - \umcu\right)\right]
\right)\,,
\eea
\bea
\label{eq:qpm}
\left\vert\left(\dfrac{q}{p}\right)_q\right\vert-1&=&
-\dfrac{{\cal K}}{2\,\langle \overline B_q \vert {\cal O}^q_1 \vert B_q
\rangle} \times\nn\\
&&\left(\dsum\limits_{i=1,2}\langle \overline B_q \vert {\cal O}^q_i \vert
B_q\rangle \left[\dfrac{\sin 2\beta_q}{R^2_{t\,q}}\,\left( D^{cc}_i
+ D^{uu}_i -2 \,D^{cu}_i\right)- 2\,\dfrac{\sin\beta_q}{R_{t\,q}}\,\left(
D^{uu}_i - D^{cu}_i \right)\right]\right.\nn\\
&&+\left. \left[\dfrac{\sin 2\beta_q}{R^2_{t\,q}}\,\left(\umcc + \umuu -2
\,\umcu\right) -2\,\dfrac{\sin\beta_q}{R_{t\,q}}\,\left( \umuu - \umcu
\right)\right]\right)\,.
\eea
The two operators entering these expansions at the leading order, ${\cal O}
^q_1$ and ${\cal O}^q_2$, are defined in eq.~(\ref{eq:effop}). The coefficient
${\cal K}$ is given by ${\cal K}=4\pi m_b^2/(3 M_W^2 \eta_{B} S_0(x_t))$.
In addition we have used
\be
V^*_{cb}V_{cq}/(V^*_{tb}V_{tq})=-e^{i\beta_q}/R_{t\,q}\, .
\ee
$\beta_d$ and $R_{t\,d}$ are the usual angle $\beta$ and the side $R_t$ of the unitarity triangle
respectively, whereas $\beta_s$ and $R_{t\,s}$ parameterize the Cabibbo-suppressed
contributions to the $B_s$ system. For completeness, we give their expansion
in terms of the parameters of CKM matrix, up to and including ${\cal O}(\lambda^4)$ terms.
They read
\be
\begin{array}{l l}
R_{td}=\sqrt{(1-\bar\rho)^2+\bar\eta^2}\,, &\sin\beta_d=\bar\eta/R_{td}\,,\cr
R_{ts}=1+\lambda^2\bar\rho+\frac{1}{2}\lambda^4(2\bar\rho+\bar\eta^2)\,, &\sin\beta_s=-\lambda^2\bar\eta
\left[1+\lambda^2\left(1-\bar\rho\right)\right]\,.
\end{array}
\ee
The Cabibbo-suppressed contributions are practically irrelevant for $\ds$ so that,
to an excellent approximation, one can put $\beta_s=0$ and $R_{t\,s}=1$ in
eq.~(\ref{eq:dgm}) to obtain
\bea
\Delta\Gamma_s &=& \dfrac{\Delta m_s \,{\cal K}}{\langle \overline B_s \vert
{\cal O}^s_1 \vert B_s\rangle}\, \left(\dsum\limits_{i=1,2}\langle \overline
B_s \vert {\cal O}^s_i \vert B_s\rangle D^{cc}_i + \usmcc +{\cal O}
(\lambda^2)\right)\,.
\eea
The contributions neglected in the previous equation, however, give rise to the
CP violation effects in the $\bsbs$ mixing, which are accounted for by the
deviation of the parameter $\vert(q/p)_s\vert$ from unity. Using $\beta_s=0$ and
$R_{t\,s}=1$ in eq.~(\ref{eq:qpm}) gives in fact $\vert(q/p)_s\vert=1$.

For the matrix elements entering our calculation (see eqs.~(\ref{eq:effop}) and
(\ref{rrt1})), we use the following parameterization
\be
\begin{array}{ll}
\langle \overline B_q\vert {\cal O}^q_1 \vert B_q\rangle=
\dfrac{8}{3}\,f_{B_q}^2 M^2_{B_q} \,B^{q}_1 \, , &
\langle \overline B_q\vert {\cal O}^q_2 \vert B_q\rangle=
-\dfrac{5}{3}\,\dfrac{f_{B_q}^2 M^4_{B_q}}{(m_b+m_q)^2} \,B^{q}_2\,,\cr
  \langle  \overline  B_q | R^q_1 | B_q \rangle  =
\dfrac{7}{3}\,\dfrac{m_q}{m_b}\, f_{B_q}^2 M_{B_q}^2\,B^{q}_{R_1},&
\langle  \overline  B_q | R^q_2 | B_q \rangle   =
 - \dfrac{2}{3}\,f_{B_q}^2 M_{B_q}^2
  \left( \dfrac{M_{B_q}^2}{m_b^2} - 1 \right)\,B^{q}_{R_2}\,,\cr
\langle  \overline  B_q | R^q_3 | B_q \rangle   =
\dfrac{7}{6}\,  f_{B_q}^2 M_{B_q}^2
  \left( \dfrac{M_{B_q}^2}{m_b^2} - 1 \right)\,B^{q}_{R_3}\,,&
\langle  \overline  B_q | R^q_4 | B_q \rangle   =
   - f_{B_q}^2 M_{B_q}^2
  \left( \dfrac{M_{B_q}^2}{m_b^2} - 1 \right)\,B^{q}_{R_4}\,.\cr
\end{array}
\label{eq:bpar1}
\ee

Among these $B$-parameters, $B^{q}_1$ and $B^{q}_2$ are the most widely
studied and well known in lattice QCD~\cite{Becirevic:2001xt},
\cite{Gimenez:2000jj}-\cite{Aoki:2003xb}.\footnote{For estimates of these matrix
elements based on QCD sum rules, see refs.~\cite{Korner:2003zk}-\cite{Narison:1994zt}.} In this paper we use the results
of ref.~\cite{Becirevic:2001xt}, in which the complete set of $\Delta B=2$,
dimension-six, four-fermion operators has been determined in the quenched
approximation of QCD. For $B^{q}_1$ and $B^{q}_2$, the results of 
\cite{Becirevic:2001xt} are in very good agreement with those obtained in
\cite{Hashimoto:2000eh,Aoki:2002bh} by using the lattice NRQCD approach. In 
addition, it has been shown in refs.~\cite{Yamada:2001xp,Aoki:2003xb} that the 
effect of the quenching approximation for these quantities is practically 
irrelevant. Thus, the systematic uncertainties in the present lattice estimates 
of $B^{q}_1$ and $B^{q}_2$ are quite under control.

To our knowledge, the matrix elements of the operators $R^q_i$ defined in
eq.~(\ref{eq:bpar1}), entering the $1/m_b$ corrections, have been only
estimated so far in the VSA.
Using the complete set of operator matrix elements calculated
in~\cite{Becirevic:2001xt}, however, two of the four independent parameters
$B^{q}_{R_i}$ can be also evaluated. Besides the operators ${\cal O}^q_{1,2}$,
the complete basis studied in~\cite{Becirevic:2001xt} includes
\be
{\cal O}^q_3=(\bar b_i q_j)_{S-P}(\bar b_j q_i)_{S-P}\,,\quad
{\cal O}^q_4=(\bar b_i q_i)_{S-P}(\bar b_j q_j)_{S+P}\,,\quad
{\cal O}^q_5=(\bar b_i q_j)_{S-P}(\bar b_j q_i)_{S+P}\,,
\label{basisc}
\ee
whose matrix elements are parametrized as
\bea
 \label{paramsdam}
\langle \overline B_q \vert {\cal O}^q_3 \vert   B_q \rangle &=& {1\over 3} \,
\left( {m_{B_q}\over m_b + m_q } \right)^2 m_{B_q}^2  f_{B_q}^2 B^q_3\, ,
\nn\\
\langle \overline B_q \vert {\cal O}^q_4 \vert   B_q \rangle &=&  2 \, \left(
{m_{B_q}\over m_b + m_q } \right)^2 m_{B_q}^2  f_{B_q}^2 B^q_4\,  ,   \\
\langle \overline B_q \vert {\cal O}^q_5 \vert   B_q \rangle  &=& {2\over 3} \,
\left( {m_{B_q}\over m_b + m_q } \right)^2 m_{B_q}^2  f_{B_q}^2 B^q_5\,  .
\nonumber
\eea
We then notice that the operator $R^q_1$ defined in eq.~(\ref{rrt1}) is
trivially related to ${\cal O}^q_4$,
\be
\label{eq:rq1}
R^q_1=\dfrac{m_q}{m_b}\,{\cal O}^q_4\,.
\ee
In addition, by using the Fierz identities and the equations of motion, the
operator $R^q_4$ can be expressed as~\cite{Beneke:1996gn}
\be
\label{eq:rq4}
2 R^q_4={\cal O}^q_3 +{\cal O}^q_1/2+{\cal O}^q_2- 2\dfrac{m_q}{m_b}\,{\cal O}^q_5 + R^q_2\,.
\ee
Thus, eqs.~(\ref{eq:rq1}) and (\ref{eq:rq4}) can be used to get rid of the two
parameters $B^{q}_{R_1}$ and $B^{q}_{R_4}$.

The values of the $B$-parameters obtained in ref.~\cite{Becirevic:2001xt} and
used in our calculation are collected in table~\ref{tab:inputs}.
Concerning the unknown matrix elements of the operators $R^q_2$ and $R^q_3$,
they have been estimated in the VSA and we have included a 30\% of relative
error to account for the corresponding systematic
uncertainty~\footnote{Since terms proportional to $m_d/m_b$ are neglected in our
calculation, the values of
$B^{d}_{4}$ and $B^{d}_{5}$ are presented in the table only for completeness.}.
\begin{table} [t]
\renewcommand{\arraystretch}{1.6}
\begin{center}
\begin{tabular}{|l|l|}
\hline
$m_b$ = $4.75\pm 0.11$ GeV & $m_c/m_b$ = $0.30\pm 0.02$  \\
${\overline m}_t({\overline m}_t)=165.0\pm 5.0$ GeV &
${\overline m}_s({\overline m}_b)=87\pm 21$ MeV\\
$M_W=80.41$ GeV & $\Delta m_d$ = $0.502\pm 0.006$ $\rm{ps}^{-1}$ \\
$\tau(B_s)=1.461\pm 0.057$ ps & $\tau(B_d)=1.540\pm 0.014$ ps \\
$M_{B_d}=5.279$ GeV & $M_{B_s}=5.369$ GeV  \\
$\eta_b(m_b)=0.85\pm 0.02$& $\xi = 1.24\pm 0.04$  \\
\hline
$\lambda=0.2240\pm 0.0036$ & $\as(m_Z)=0.118$\\
$\bar \rho= 0.162\pm 0.046$ & $\bar \eta=0.347 \pm 0.027$\\
\hline
$ B_1^s$ = $0.87 \pm 0.05$ & $ B_1^s/B_1^d$ = $0.99 \pm 0.02$ \\
$ B_2^s$ = $0.84 \pm 0.04$ & $ B_2^s/B_2^d$ = $1.01 \pm 0.02$\\
$ B_3^s$ = $0.91 \pm 0.08$ & $ B_3^s/B_3^d$ = $1.01 \pm 0.03$ \\
$ B_4^s$ = $1.16 \pm 0.07$ & $ B_4^s/B_4^d$ = $1.01 \pm 0.02$  \\
$ B_5^s$ = $1.75 \pm 0.20$ & $ B_5^s/B_5^d$ = $1.01 \pm 0.03$\\
\hline
\end{tabular}
\end{center}
\caption{\it Central values and standard deviations
of the input parameters used to obtain the theoretical
estimates of the width differences and of the CP violation parameters. When the
error is not quoted, the parameter has been kept fixed in the numerical
analysis. The values of $m_b$ and $m_c$ refer to the pole masses,
while ${\overline m}_s$ and
${\overline m}_t$ are the masses in the $\msb$ scheme. The $B$-parameters are
renormalized in the $\msb$ scheme at the scale $m_b$. The definition of the
renormalization scheme can be found in \cite{Beneke:1998sy} for $B_1^q$, $B_2^q$
and $B_3^q$ and in \cite{Misiak:1999yg} for $B_4^q$ and $B_5^q$, see also
appendix B.}
\label{tab:inputs}
\end{table}

In order to obtain the theoretical predictions presented in this paper, we have
performed a Bayesian statistical analysis by implementing a simple Monte
Carlo calculation. The input parameters have been extracted with flat
distributions, by assuming the central values and standard deviations given in
table~\ref{tab:inputs}.

\begin{figure}[t]
\begin{center}
\includegraphics[width=\textwidth]{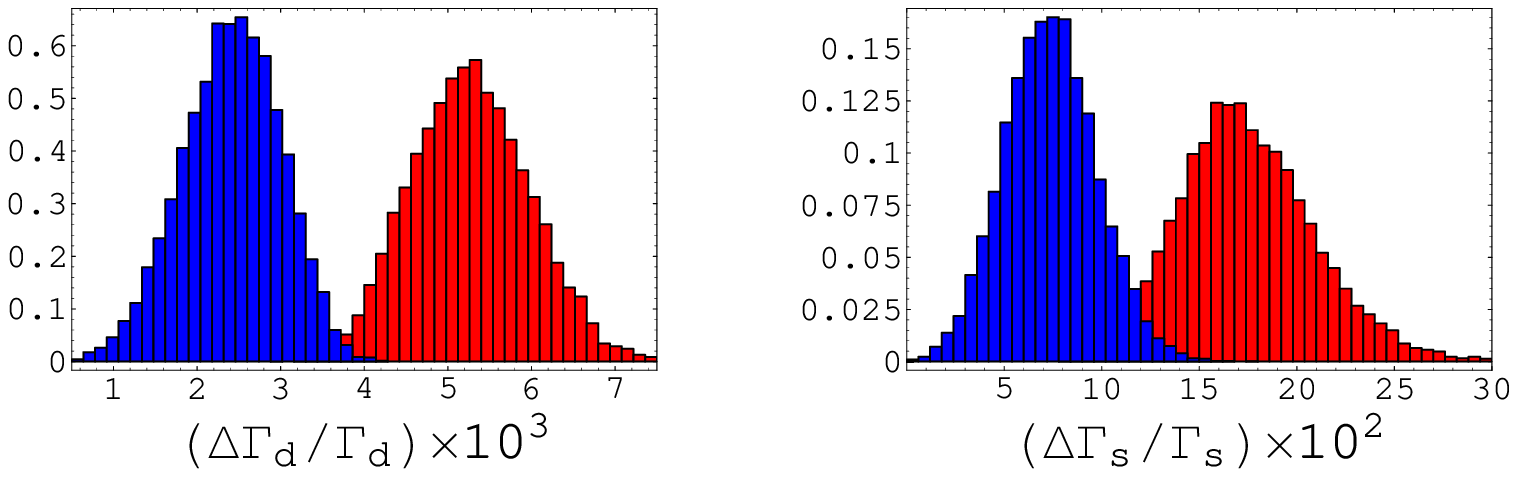}
\end{center}
\caption{\it Theoretical distributions for the width differences in the $B_d$
and $B_s$ systems. The predictions are shown at both the LO (light/red) and NLO
(dark/blue).}
\label{fig:plot1}
\end{figure}
\subsection{Width differences: \boldmath$\dgd$, \boldmath$\dgs$ and
\boldmath$\dd/\ds$}
The theoretical predictions for the width differences $\dgd$ and $\dgs$, as
obtained from eq.~(\ref{eq:dgm}) as functions of the CKM matrix elements and
of the $B$-parameters entering at the LO in $1/m_b$, can be expressed as
\bea
\label{eq:magicdgd}
\left(\frac{\Delta \Gamma_d}{\Gamma_d}\right)& = &10^{-3}\,\left[0.96(10) +
5.48(34) \dfrac{B^d_2}{B^d_1 }  -2.9(1.3)\dfrac{1}{B^d_1}\right. \nn\\
&&\phantom{10^{-3}} -\dfrac{\cos\beta }{R_t} \left(0.80(13) +
0.24(7)
\dfrac{B^d_2}{B^1_d} -0.03(6) \dfrac{1}{B^d_1} \right) \nn\\
&&\phantom{10^{-3}}\left. +\frac{\cos2\beta }{R_t^2} \left(-0.005(4) - 0.007(16)
\frac{B^d_2}{B^d_1 }  -0.008(17) \frac{1}{B^d_1} \right)\right]\,,
\eea
\bea
\label{eq:magicdgs}
\left(\frac{\Delta \Gamma_s}{\Gamma_s }\right) & = &
\frac{10^{-2}}{(\lambda\,R_{t})^2}\, \left(0.02(2) + 0.77(7) \dfrac{B^s_2}
{B^s_1 }  -0.47(18)\dfrac{1}{B^s_1}\right).
\eea
These formulae are one of the main results of this paper. In the above expressions, the
terms proportional to $1/B^q_1$ represent the contributions of the $1/m_b$
corrections. Note also that, in order to obtain the prediction for $\dgs$, the
mass difference $\Delta m_s$ has been evaluated in terms of $\Delta m_d$ by
using
\be
\Delta m_s=\Delta m_d\,\dfrac{M_{B_s}}{M_{B_d}}\,\left\vert\dfrac{V_{ts}}
{V_{td}}\right\vert^2\xi^2 \,,
\ee
where $\xi=\Big(f_{B_s}\sqrt{B^s_1}\Big)/\Big(f_{B_d}\sqrt{B^d_1}\Big)$.

The errors on the numerical coefficients presented in eqs.~(\ref{eq:magicdgd})
and (\ref{eq:magicdgs}) take into account both the residual NNLO dependence on
the renormalization scale of the $\DB=1$ operators and the theoretical
uncertainties on the various input parameters. To estimate the former, the
scale $\mu_1$ has been varied in the interval between $m_b/2$ and $2 m_b$.
These errors are strongly correlated, since they originate from the theoretical
uncertainties on the same set of input parameters. For this reason, they have
not been used to derive our final predictions for the width differences. For
these predictions, we quote the average and the standard deviation of the
corresponding probability distribution functions obtained directly from the
Monte Carlo simulation, namely
\be
\label{eq:nlores2}
\dgd=\Rdggd\, ,  \qquad
\dgs=\Rdggs\, .
\ee

The theoretical distributions are shown in fig.~\ref{fig:plot1}. Notice that the
difference between the NLO and LO distributions is remarkable. The NLO 
corrections decrease the values of both $\dgd$ and $\dgs$ by about a factor two 
with respect to the LO predictions.
We find that the total error on the width differences is dominated by the uncertainty on
the $b$-quark mass in the first place, followed closely by the one due to
the renormalization scale variation. Other contributions to the error, coming from
the ratio $m_c/m_b$, the $B$- and the CKM-parameters are smaller, though not entirely
negligible.

\begin{figure}[t]
\begin{center}
\includegraphics[width=0.7\textwidth]{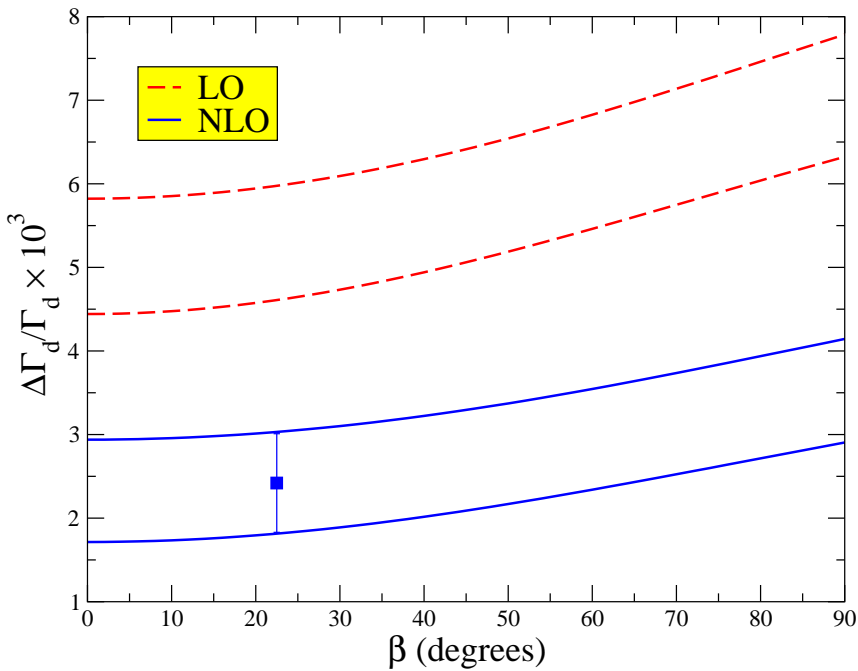}
\end{center}
\caption{\it
Prediction for $\dgd$ as a function of $\beta$ as obtained at the LO and NLO.
The NLO estimate of $\dgd$ at the measured value of $\beta$ is also shown.}
\label{fig:dgdvbeta}
\end{figure}
In fig.~\ref{fig:dgdvbeta} we show the dependence of $\dgd$ on the mixing phase
$\beta$ as given in eq.~(\ref{eq:magicdgd}). NLO corrections in the limit
of vanishing $m_c$ just shift the LO curve, while $m_c$-dependent terms
modify also its profile. In principle, a measurement of $\dgd$ could allow a
determination of $\beta$, but we find that the dependence is so mild that the extracted
value would be affected by a very large error.

Another interesting prediction concerns the ratio $\dd/\ds$. This quantity is of
particular interest, because the uncertainties coming from higher order QCD and
$\Lambda_{QCD}/m_b$ corrections, as well as those coming from the
non-perturbative estimates of the $B$-parameters, are expected to cancel in this
ratio to some extent. From our numerical analysis, we obtain the NLO prediction
\be
\dd/\ds=\Rdgdgs\,.
\ee
\begin{figure}[t]
\begin{center}
\includegraphics[width=0.55\textwidth]{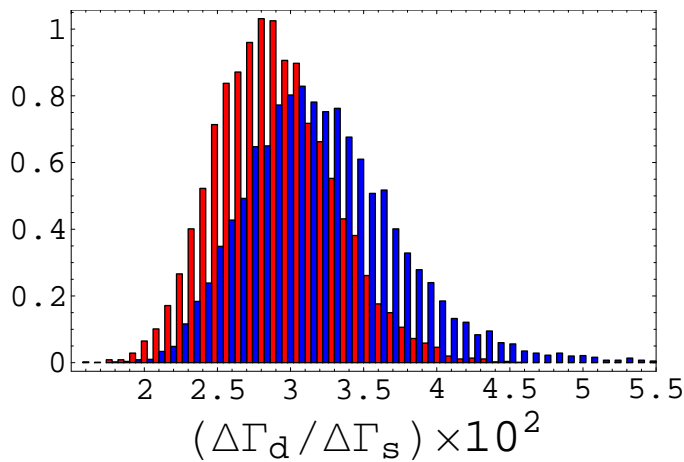}
\end{center}
\caption{\it
Theoretical distribution for the ratio $\dd/\ds$ as obtained at both the LO
(light/red) and NLO (dark/blue).}
\label{fig:plot2}
\end{figure}
The corresponding theoretical distributions at the LO and NLO are shown in fig.~\ref{fig:plot2}. As
can be seen from the plot, this quantity is practically unaffected by the NLO
corrections.

\subsection{CP Violation parameters: \boldmath$\vert(q/p)_d\vert$ and
\boldmath$\vert(q/p)_s\vert$}
CP violation in the mixing shows up only for a non-vanishing value of the charm
quark mass, as can be explicitly verified from the expression of $\vert(q/p)_q
\vert$ in eq.~(\ref{eq:qpm}). In addition, it is apparent from
eq.~(\ref{eq:qpm}) that $\vert(q/p)_s\vert-1$ is suppressed with respect to
$\vert(q/p)_d\vert-1$ by a factor $\sin\beta_s\sim\lambda^2$.

For the parameters $\vert(q/p)_d\vert$ and $\vert(q/p)_s\vert$, expressed
in terms of the CKM matrix elements and the $B$-parameters $B^q_1,\,B^q_2$, we
obtain the expression
\bea
\label{eq:magicpqq}
\left\vert\left(\dfrac{q}{p}\right)_q\right\vert-1& = & 10^{-4}\left[
\dfrac{\sin\beta_q }{R_{tq}}
\left(5.19(84) + 1.57(47) \dfrac{B^q_2}{B^q_1}-0.22(39) \dfrac{1}{B^q_1} \right)
\right.\nn\\
&&\phantom{10^{-4}} \left.+\frac{\sin 2\beta_q }{R_{tq}^2}
\left(0.04(3) + 0.04(10)\frac{B^q_2}{B^q_1 }+0.05(11) \frac{1}{B^q_1}\right)
\right]\,,
\eea
where $q=d,s$.

\begin{figure}[t]
\begin{center}
\includegraphics[width=\textwidth]{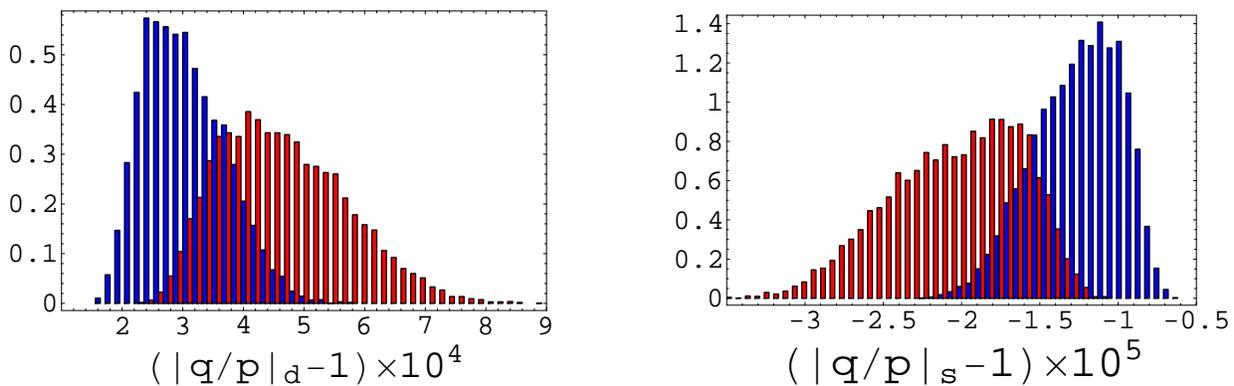}
\end{center}
\caption{\it Theoretical distributions for $\vert(q/p)_q\vert-1$ in the $B_d$
and $B_s$ systems. The predictions are shown at the LO (light/red) and
NLO (dark/blue).}
\label{fig:plot3}
\end{figure}

By using the values of the CKM and $B$-parameters given in
table~\ref{tab:inputs} we obtain from the Monte Carlo analysis the final
predictions
\be
\label{eq:nlores3}
\left\vert\left(\dfrac{q}{p}\right)_d\right\vert-1=\Rqpd\quad,\quad
\left\vert\left(\dfrac{q}{p}\right)_s\right\vert-1=\Rqps \,.
\ee
The theoretical distributions for these quantities are shown in
fig.~\ref{fig:plot3} both at the LO and NLO.
One can see that the effect of the NLO corrections
turns out to be important also for these quantities. The error on
$\vert(q/p)_q\vert-1$ is largely dominated by the uncertainties on $m_c/m_b$
and on the CKM parameters. We note in particular that the dependence on
the $B$-parameters is practically negligible, making
these predictions free from hadronic uncertainties.

\begin{figure}[t]
\begin{center}
\includegraphics[width=0.7\textwidth]{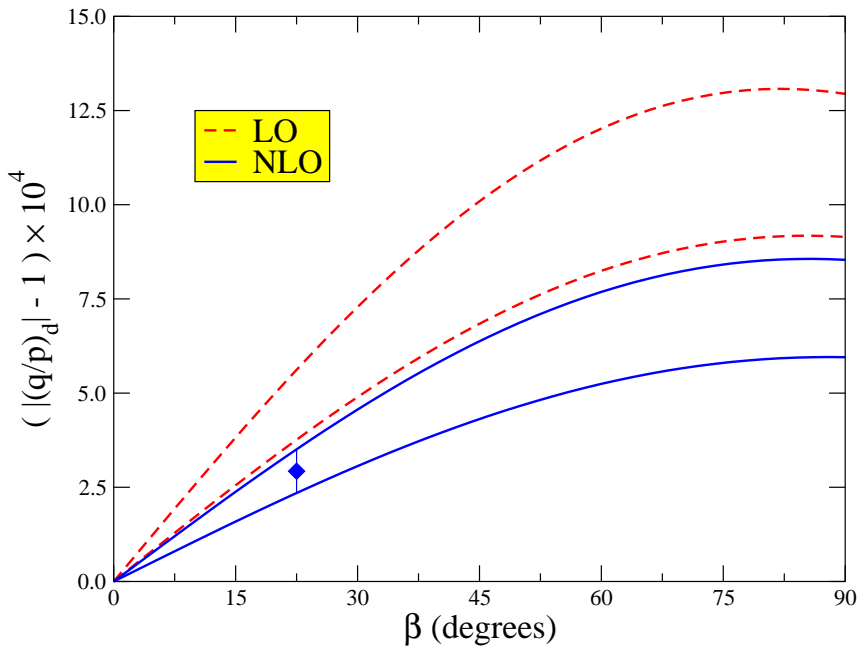}
\end{center}
\caption{\it
Prediction for the CP violation parameter $\vert(q/p)_d\vert-1$
as a function of $\beta$ as obtained at both the LO and NLO. The NLO estimate of
$\vert(q/p)_d\vert-1$ at the measured value of $\beta$ is also shown.}
\label{fig:qpdvbeta}
\end{figure}
Finally,  we show in fig.~\ref{fig:qpdvbeta} the LO and NLO predictions for $\vert(q/p)_d
\vert-1$ as functions of $\beta$. In this case, NLO corrections without charm mass leave the LO
result practically unchanged, while the $m_c$-dependent terms give a sizable contribution.
The dependence on $\beta$ is quite strong, allowing for a determination of the mixing angle once
$\vert(q/p)_d\vert$ will be measured.

\section{Conclusions}
The main result of this paper is the NLO QCD calculation of the Wilson
coefficients entering the heavy quark expansion of
$\Gamma_{21}^d$ and $\Gamma_{21}^s$, including the effects
of a non-vanishing charm quark mass. For $\Gamma_{21}^s$ we find agreement with
a previous result~\cite{Beneke:1998sy}. Using our results, we improve
the theoretical predictions for different observables in the neutral $B$-meson
systems, namely the width differences $\dgd$ and $\dgs$ and the CP violation
parameters $\vert (q/p)_d\vert$ and $\vert (q/p)_s\vert$.

Using formulae including the NLO $\alpha_s$ corrections at the lowest order in the
$1/m_b$ expansion and the first power corrections at the LO in $\alpha_s$, we find
$$
\frac{\dd}{\Gamma_d}=\Rdggd\,,\quad\frac{\ds}{\Gamma_s}=\Rdggs\,,\quad   \frac{\dd}{\ds}= \Rdgdgs\,,
$$
\be
\left\vert\left(q/p\right)_d\right\vert-1=\Rqpd\,,\quad
\left\vert\left(q/p\right)_s\right\vert-1=\Rqps\,.
\ee

NLO QCD corrections are in general important for theoretical consistency and give
sizable contributions to the Wilson coefficients considered in this paper.
We find that the charm quark mass effects at NLO are numerically important, in particular
for $\vert (q/p)_q\vert$, and should be included in the phenomenological analyses.

\section*{Acknowledgments}
We thank D.~Becirevic and J.~Reyes for interesting discussions and suggestions
on the subject of this paper. We also thank F.~Martinez for correspondence on
the experimental issues.
Work partially supported by the European
Community's Human Potential Programme under HPRN-CT-2000-00145 Hadrons/Lattice
QCD.

\section*{Note added}
Just before this paper was finalized, a paper by Beneke {\it et al.} on the
same subject was submitted to the e-print archive~\cite{Beneke:2003az}.
Comparing the results, we found that the NLO contribution to the Wilson
coefficients $D_k^{cu}$ of
eqs.~(\ref{eq:dccs}) and (\ref{eq:c6db0}) differ from the ones given in eq.~(25) of
ref.~\cite{Beneke:2003az}.
Had we assumed that the diagrams $D_6$, $D_7$ and $D_8$ in fig.~\ref{fig:fddb1}
were equal to the corresponding ``rotated'' ones ({\it i.e.}~those diagrams where the gluon
is attached to the other external quark leg with the same flavour and to the other virtual
quark line), we would have obtained
the result of ref.~\cite{Beneke:2003az}. However this assumption, which is correct
when the two virtual lines correspond to the same flavour, does not hold when the two
quarks are different. This may explain the origin of the discrepancy.

\section*{Appendix A: Analytical Results for the \boldmath$\DB=2$ Wilson
Coefficients}
\label{sec:Anares}

In this appendix we collect the analytical expressions of the Wilson
coefficients, both at the LO and NLO. The LO coefficients have been computed
in refs.~\cite{Beneke:1996gn,Dighe:2001gc} and are given here for completeness.

The coefficient functions $D^{uu}_k$, $D^{cu}_k$ and $D^{cc}_k$ defined in 
eq.(\ref{eq:dccs}) depend quadratically on the coefficient functions $C_i$ of 
the $\DB=1$ effective Hamiltonian and can be written as
\bea
D^{uu}_k(\mu_2) &=& \dsum_{i,j=1,2} C_i(\mu_1) \, C_j(\mu_1) F^{uu}_{k,ij}
(\mu_1,\mu_2)+\frac{\as}{4\pi}\,C_2(\mu_1)^2 P^{uu}_{k,22}(\mu_1,\mu_2)
\nn\\
&+& 2 \,\frac{\as}{4\pi} \, C_2 \, C_{8G} P^{u}_{k,28}+2\, \dsum_{i=1,2} 
\dsum_{r=3,6} C_i \, C_r P^{u}_{k,ir}\,,
\nn \\
D^{cu}_k(\mu_2) &=& \sum_{i,j=1,2} C_i(\mu_1)\, C_j(\mu_1)\, F^{cu}_{k,ij}
(\mu_1,\mu_2) +\frac{\as}{4\pi}\,C_2(\mu_1)^2 P^{cu}_{k,22}(\mu_1,\mu_2)
\nn\\
&+& \frac{\as}{4\pi} \,C_2 \, C_{8G} \left(P^{c}_{k,28}+P^{u}_{k,28}\right)
+ \,\dsum_{i=1,2} \dsum_{r=3,6} C_i \, C_r \left( P^{c}_{k,ir} +P^{u}_{k,ir}
\right)\,,\nn\\
D^{cc}_k(\mu_2) &=& \dsum_{i,j=1,2} C_i(\mu_1) \, C_j(\mu_1) F^{cc}_{k,ij}
(\mu_1,\mu_2)+\frac{\as}{4\pi}\,C_2(\mu_1)^2 P^{cc}_{k,22}(\mu_1,\mu_2)
\nn\\
&+& 2 \,\frac{\as}{4\pi} \, C_2 \, C_{8G} P^{c}_{k,28}+2\, \dsum_{i=1,2}
\dsum_{r=3,6} C_i \, C_r P^{c}_{k,ir}\,.
\label{eq:c6db0}
\eea
The functions $F^{qq^\prime}_{k,ij}$ are obtained from the insertion of the
operators $Q_i$ and $Q_j$ in the Feynman diagrams $D_1-D_{10}$ of fig.~\ref{fig:fddb1},
whereas the
diagrams $D_{11}$ and $D_{12}$ contribute to the functions $P^{qq^\prime}_{k,22}
(\mu_1,\mu_2)$ and $P^{q}_{k,ij}$ respectively. Contributions with the double
insertion of penguin operators have been neglected, since the Wilson coefficients
$C_3$--$C_6$ are numerically small.

We distinguish the leading and next-to-leading contributions in the coefficients
$F^{qq^\prime}_{k,ij}$ by writing the expansion
\be
F^{qq^\prime}_{k,ij} = A^{q q^\prime}_{k,ij} + \frac{\as}{4\pi} B^{q q^\prime}_
{k,ij}\,,
\ee
where $(q q^\prime)=\{(uu),(cu),(cc)\}$. Since  only the
sum of $F^{qq^\prime}_{k,12}$ and $F^{qq^\prime}_{k,21}$ contributes to eq.~(\ref{eq:c6db0}),
we just
give in the following the average of the components 12 and 21. Moreover, we do
not write explicitly the results for the functions $D^{uu}_k$ since they can be
obtained by taking the limit $m_c\to 0$ of $D^{cc}_k$.

In terms of the ratio $z=m_c^2/m_b^2$, the LO coefficients $A^{qq^\prime}_
{k,ij}$ read
\begin{footnotesize}
\be
\begin{array}{lll}
  A^{cu}_{1,11}={\frac{3}{2}}\,\left( 2 - 3\,z + {z^3} \right) \,, \qquad
& A^{cu}_{1,12}=A^{cu}_{1,21}={\frac{1}{2}}\, \left(2 - 3\,z + {z^3}\right)\,,\qquad
& A^{cu}_{1,22}={\frac{1}{2}}\,{\left( 1 - z \right) }^3\,,\\
  A^{cu}_{2,11}=3\,{{\left( 1 - z \right) }^2}\,\left( 1 + 2\,z \right)\,, \qquad
& A^{cu}_{2,12}=A^{cu}_{2,21}={{\left( 1 - z \right) }^2}\,\left( 1 + 2\,z \right)\,, \qquad
& A^{cu}_{2,22}=- {{\left( 1 - z \right) }^2}\,\left( 1 + 2\,z \right) \,.
\end{array}
\ee
\be
\begin{array}{lll}
  A^{cc}_{1,11}=3\,{\sqrt{1 - 4\,z}}\,\left( 1 - z \right)\,, \qquad
& A^{cc}_{1,12}={\sqrt{1 - 4\,z}}\,\left( 1 - z \right)\,,\qquad
& A^{cc}_{1,22}={\frac{1}{2}}\,{\left( 1 - 4\,z \right) }^{\frac{3}{2}}\,,\\
  A^{cc}_{2,11}=3\,{\sqrt{1 - 4\,z}}\,\left( 1 + 2\,z \right)\,, \qquad
& A^{cc}_{2,12}={\sqrt{1 - 4\,z}}\,\left( 1 + 2\,z \right)\,, \qquad
& A^{cc}_{2,22}=- {\sqrt{1 - 4\,z}}\,\left( 1 + 2\,z \right) \,. 
\end{array}
\ee
\end{footnotesize}
\renewcommand{\arraystretch}{2.0} 

The NLO results for the coefficients $B^{qq^\prime}_{k,ij}$ are presented in the
$\msb$ scheme of ref.~\cite{db2nlo1} for the $\DB=1$ operators and the $\msb$ 
scheme of ref.~\cite{Beneke:1998sy} for the $\DB=2$ operators, in QCD. We find

\renewcommand{\arraystretch}{2.5} 
\begin{footnotesize}

\bea
B^{cu}_{1,11}&=&{\frac{109}{6}} - 37\,z + {\frac{3\,{z^2}}{2}} + {\frac{52\,{z^3}}{3}} +
  2\,{{\left( 1 - z \right) }^2}\,\left( 5 + z \right) \,
    \log x_2 - 4\,{{\left( 1 - z \right) }^2}\,
   \left( 5 + 7\,z \right) \,\log (1 - z) - \nn\\
& &  2\,z\,\left( 10 + 14\,z - 15\,{z^2} \right) \,\log z + 
  8\,\left( 2 - 3\,z + {z^3} \right) \,\log (1 - z)\,\log z +16\,\left( 2 - 3\,z + {z^3} \right) \,\mathrm{Li_2}(z)\,,\nn\\
B^{cu}_{2,11}&=&{-\frac{4}{3}\,\left( 10 - 33\,z + 54\,{z^2} - 31\,{z^3} \right) } - 
  8\,{{\left( 1 - z \right) }^2}\,\left( 4 + 14\,z - 3\,{z^2} \right) \,
   \log (1 - z) +\nn\\
& & 8\,z\,\left( 2 - 23\,z + 21\,{z^2} - 3\,{z^3} \right) \,
   \log z -\nn\\
& & 16\,{{\left( 1 - z \right) }^2}\,\left( 1 + 2\,z \right) \,
   \left( 2\, \log x_2 - \log (1 - z)\,\log z - 
     2\,\mathrm{Li_2}(z) \right)\,,
\eea

\bea
\dfrac{B^{cu}_{1,12}+B^{cu}_{1,21}}{2}&=& {\frac{-502 + 912\,z - 387\,{z^2} - 23\,{z^3}}{36}} - 
  {{\left( 1 - z \right) }^2}\,\left( 17 + 4\,z \right) \,
   \log x_1 + \frac{2}{3}\left( 1 - z \right)^2\,
      \left( 5 + z \right) \, \log x_2 - \nn\\
& &  {\frac{{{\left( 1 - z \right)}^2}}{12\,z}\,\left( 2 + 33\,z + 94\,{z^2} \right) \,
      \log (1 - z)} -{\frac{z}{12}
      \left( 80 + 69\,z - 126\,{z^2} \right) \,\log z} +\nn\\ 
& &  {\frac{8}{3}\,\left( 2 - 3\,z + {z^3} \right) \,
      \left( \log (1 - z)\,\log z + 2\,\mathrm{Li_2}(z) \right)
    }\,,\nn\\
\dfrac{B^{cu}_{2,12}+B^{cu}_{2,21}}{2}&=&{\frac{-130 + 93\,z + 144\,{z^2} - 107\,{z^3}}{9}} - 
  {\frac{2\,{{\left( 1 - z \right) }^2}}{3\,z}\,
      \left( 1 + 15\,z + 47\,{z^2} - 12\,{z^3} \right) \,\log (1 - z)}
   +\nn\\
& &{\frac{2}{3}\,z\,\left( 8 - 93\,z + 87\,{z^2} - 12\,{z^3} \right) \,\log z
    } -\nn\\
& & {\frac{8}{3}\,{{\left( 1 - z \right) }^2}\,\left( 1 + 2\,z \right) \,
      \left( 3\,\log x_1 + 4\, \log x_2 - 
        2\,\log (1 - z)\,\log z - 4\,\mathrm{Li_2}(z) \right)}\,,\eea

\bea
B^{cu}_{1,22}&=& {-\frac{{{\pi }^2}}{3}\,\left( 1 - 5\,z + 4\,{z^2} \right)
    } + {\frac{-136 - 159\,z + 738\,{z^2} - 443\,{z^3}}{18}} - 
  2\,{{\left( 1 - z \right) }^2}\,\left( 5 + 4\,z \right) \,
   \log x_1 +\nn\\
& &  {\frac{2}{3}\,{{\left( 1 - z \right) }^2}\,
      \left( 4 - z \right) \, \log x_2} + 
  {\frac{{{\left( 1 - z \right) }^2}}{6\,z}\,
      \left( 7 + 32\,{z^2} + 3\,{z^3} \right) \,\log (1 - z)} -\nn\\
& & {\frac{z}{6}\,\left( 62 - 39\,z - 30\,{z^2} + 3\,{z^3} \right) \,\log z
    } +{\frac{\left( 5 - 3\,z - 18\,{z^2} + 16\,{z^3} \right)}{3} \,
      \left( \log (1 - z)\,\log z + 2\,\mathrm{Li_2}(z) \right)
    }\,,\nn\\
B^{cu}_{2,22}&=&{\frac{8}{3}\,{{\pi }^2}\,\left( 1 + z - 2\,{z^2} \right)} - 
  {\frac{28}{9}\,\left( 5 + 3\,z - 27\,{z^2} + 19\,{z^3} \right)} - 
  16\,{{\left( 1 - z \right) }^2}\,\left( 1 + 2\,z \right) \,
   \log x_1 +\nn\\
& & {\frac{32}{3}\,{{\left( 1 - z \right) }^2}\,
      \left( 1 + 2\,z \right) \, \log x_2} -
  {\frac{4\,{{\left( 1 - z \right) }^2}}{3\,z}\,
      \left( 1 - 12\,z - 16\,{z^2} - 3\,{z^3} \right) \,\log (1 - z)} +\nn\\
& &{\frac{4}{3}\,z\,\left( 2 - 3\,z + 18\,{z^2} - 3\,{z^3} \right) \,\log z}
     +{\frac{8}{3}\,\left( 1 - 3\,z - 6\,{z^2} + 8\,{z^3} \right) \,
      \left( \log (1 - z)\,\log z + 2\,\mathrm{Li_2}(z) \right) 
    }   \,,
\eea

\bea
B^{cc}_{1,11}&=& {\frac{{\sqrt{1 - 4\,z}}\,\left( 109 - 226\,z + 168\,{z^2} \right) }{6}} - 
  \left( 52 - 104\,z - 16\,{z^2} + 56\,{z^3} \right) \,\log \sigma + \nn\\
& &  2\,\left( 5 - 8\,z \right) \,{\sqrt{1 - 4\,z}}\, \log x_2 -  12\,{\sqrt{1 - 4\,z}}\,\left( 3 - 2\,z \right) \,\log (1 - 4\,z) + 
  4\,\left( 13 - 10\,z \right) \,{\sqrt{1 - 4\,z}}\,\log z +\nn\\
& & 16\,\left( 1 - 3\,z + 2\,{z^2} \right) \,
   \left( 3\,{{\log}^2 \sigma} + 2\,\log \sigma\,\log (1 - 4\,z) - 
     3\,\log \sigma\,\log z + 4\,\mathrm{Li_2}(\sigma) + 
     2\,\mathrm{Li_2}({\sigma^2}) \right)\,,\nn\\
B^{cc}_{2,11}&=&{\frac{-8\,{\sqrt{1 - 4\,z}}\,\left( 5 - 23\,z - 42\,{z^2} \right) }{3}} -
  16\,\left( 4 - 2\,z - 7\,{z^2} + 14\,{z^3} \right) \,\log \sigma - 
  32\,{\sqrt{1 - 4\,z}}\,\left( 1 + 2\,z \right) \, \log x_2 -\nn\\ 
& &  48\,{\sqrt{1 - 4\,z}}\,\left( 1 + 2\,z \right) \,\log (1 - 4\,z) + 
  64\,{\sqrt{1 - 4\,z}}\,\left( 1 + 2\,z \right) \,\log z +\nn\\
& & 16\,\left( 1 - 4\,{z^2} \right) \,
   \left( 3\,{{\log}^2 \sigma} + 2\,\log \sigma\,\log (1 - 4\,z) - 
     3\,\log \sigma\,\log z + 4\,\mathrm{Li_2}(\sigma) + 
     2\,\mathrm{Li_2}({\sigma^2}) \right)\,,
\eea

\bea
B^{cc}_{1,12}&=& {\frac{- {\sqrt{1 - 4\,z}}\,
        \left( 127 - 199\,z - 75\,{z^2} \right)   }{9}} + 
  {\frac{\left( 2 - 259\,z + 662\,{z^2} - 76\,{z^3} - 200\,{z^4} \right) \,
      \log \sigma}{12\,z}} -\nn\\
& & \left( 17 - 26\,z \right) \,{\sqrt{1 - 4\,z}}\,
   \log x_1 + {\frac{2\,\left( 5 - 8\,z \right) \,
      {\sqrt{1 - 4\,z}}\, \log x_2}{3}} - 
  4\,{\sqrt{1 - 4\,z}}\,\left( 3 - 2\,z \right) \,\log (1 - 4\,z) -\nn\\
& & {\frac{{\sqrt{1 - 4\,z}}\,\left( 2 - 255\,z + 316\,{z^2} \right) \,\log z}
    {12\,z}} +\nn\\
& & {\frac{16\,\left( 1 - 3\,z + 2\,{z^2} \right) \,
      \left( 3\,{{\log}^2 \sigma} + 2\,\log \sigma\,\log (1 - 4\,z) - 
        3\,\log \sigma\,\log z + 4\,\mathrm{Li_2}(\sigma) + 
        2\,\mathrm{Li_2}({\sigma^2}) \right) }{3}}\,,\nn\\
B^{cc}_{2,12}&=&{\frac{-2\,{\sqrt{1 - 4\,z}}\,\left( 68 + 49\,z - 150\,{z^2} \right) }{9}} + 
  {\frac{2\,\left( 1 - 35\,z + 4\,{z^2} + 76\,{z^3} - 100\,{z^4} \right) \,
      \log \sigma}{3\,z}} +\nn\\
& &  \left( 16 - 64\,{z^2} \right) \,{{\log}^2 \sigma} - 8\,{\sqrt{1 - 4\,z}}\,\left( 1 + 2\,z \right) \,\log x_1 -
  {\frac{32\,{\sqrt{1 - 4\,z}}\,\left( 1 + 2\,z \right) \,
       \log x_2}{3}} - \nn\\
& &   16\,{\sqrt{1 - 4\,z}}\,\left( 1 + 2\,z \right) \,\log (1 - 4\,z) -{\frac{2\,{\sqrt{1 - 4\,z}}\,\left( 1 - 33\,z - 76\,{z^2} \right) \,
      \log z}{3\,z}} +\nn\\
& &{\frac{16\,\left( 1 - 4\,{z^2} \right) \,
      \left( 2\,\log \sigma\,\log (1 - 4\,z) - 3\,\log \sigma\,\log z + 
        4\,\mathrm{Li_2}(\sigma) + 2\,\mathrm{Li_2}({\sigma^2}) \right)
      }{3}}\,,
\eea

\bea
B^{cc}_{1,22}&=& {\frac{- {{\pi }^2}\,\left( 1 - 10\,z \right)   }{3}} -
  {\frac{{\sqrt{1 - 4\,z}}\,\left( 115 + 632\,z + 96\,{z^2} \right) }{18}} -
  {\frac{\left( 7 + 13\,z - 194\,{z^2} + 304\,{z^3} - 64\,{z^4} \right) \,
      \log \sigma}{6\,z}} -\nn\\
& & 2\,{\sqrt{1 - 4\,z}}\,\left( 5 - 2\,z \right) \,
   \log x_1 + {\frac{4\,\left( 2 - 5\,z \right) \,
      {\sqrt{1 - 4\,z}}\, \log x_2}{3}} -
  4\,\left( 1 - 6\,z \right) \,{\sqrt{1 - 4\,z}}\,\log (1 - 4\,z) + \nn\\
& &  {\frac{\left( 13 - 54\,z + 8\,{z^2} \right) \,\log \sigma\,\log (1 - 4\,z)}
    {3}} + {\frac{{\sqrt{1 - 4\,z}}\,\left( 7 + 27\,z - 250\,{z^2} \right) \,
      \log z}{6\,z}} +\nn\\
& &\left( 7 - 32\,z + 4\,{z^2} \right) \,
   \left( {{\log}^2 \sigma} - \log \sigma\,\log z \right)  + {\frac{4\,\left( 5 - 12\,z + 4\,{z^2} \right) \,\mathrm{Li_2}(\sigma)}
    {3}} + {\frac{4\,\left( 4 - 21\,z + 2\,{z^2} \right) \,
      \mathrm{Li_2}({\sigma^2})}{3}}\,,\nn\\
B^{cc}_{2,22}&=&{\frac{8\,{{\pi }^2}\,\left( 1 + 2\,z \right) }{3}} -
  {\frac{8\,{\sqrt{1 - 4\,z}}\,\left( 19 + 53\,z + 24\,{z^2} \right) }{9}} +
  {\frac{4\,\left( 1 + 7\,z + 10\,{z^2} - 68\,{z^3} + 32\,{z^4} \right) \,
      \log \sigma}{3\,z}} -\nn\\
& & 8\,{{\left( 1 + 2\,z \right) }^2}\,{{\log}^2 \sigma} -
  16\,{\sqrt{1 - 4\,z}}\,\left( 1 + 2\,z \right) \,\log x_1 +
  {\frac{32\,{\sqrt{1 - 4\,z}}\,\left( 1 + 2\,z \right) \,
       \log x_2}{3}} + \nn\\
& &  16\,{\sqrt{1 - 4\,z}}\,\left( 1 + 2\,z \right) \,\log (1 - 4\,z) -{\frac{8\,\left( 1 + 6\,z + 8\,{z^2} \right) \,\log \sigma\,\log (1 - 4\,z)}
    {3}} -\nn\\
& & {\frac{4\,{\sqrt{1 - 4\,z}}\,\left( 1 + 9\,z + 26\,{z^2} \right) \,
      \log z}{3\,z}} +8\,{{\left( 1 + 2\,z \right) }^2}\,\log \sigma\,
   \log z +\nn\\
& & {\frac{32\,\left( 1 - 4\,{z^2} \right) \,
      \mathrm{Li_2}(\sigma)}{3}} -
  {\frac{32\,\left( 1 + 3\,z + 2\,{z^2} \right) \,
      \mathrm{Li_2}({\sigma^2})}{3}}\,,
\eea
\end{footnotesize}
\renewcommand{\arraystretch}{1.0}
\noindent where $\sigma$ is the ratio
\be
\sigma = \frac{1-\sqrt{1-4z}}{1+\sqrt{1-4z}} \, ,
\ee
and we have defined $x_1=\mu_1/m_b$ and $x_2=\mu_2/m_b$.

The contributions of the diagram $D_{11}$ in fig.~\ref{fig:fddb1} and of the 
insertions of the penguin and chromomagnetic operators, read
\begin{footnotesize}

\bea
P^{cu}_{1,22}&=&-{\frac{1}{27}} - {\frac{2\,z}{9}} - {\frac{\log x_1}{9}} -
  {\frac{{\sqrt{1 - 4\,z}}\,\left( 1 + 2\,z \right) \,
      \left( 2 + 3\,\log \sigma + 6\,\log x_1 \right) }{54}} +
  {\frac{\log z}{18}}\,,\nn\\
P^{cu}_{2,22}&=&{\frac{8}{27}} + {\frac{16\,z}{9}} + {\frac{8\,\log x_1}{9}} +
  {\frac{4\,{\sqrt{1 - 4\,z}}\,\left( 1 + 2\,z \right) \,
      \left( 2 + 3\,\log \sigma + 6\,\log x_1 \right) }{27}} -
  {\frac{4\,\log z}{9}}\,,
\eea

\bea
P^{cc}_{1,22}&=&{\frac{-2\,{\sqrt{1 - 4\,z}}\,\left( 1 + 8\,z + 12\,{z^2} \right) }{27}} -
  {\frac{\log \sigma}{9}} + {\frac{4\,{z^2}\,\log \sigma}{3}} +
  {\frac{16\,{z^3}\,\log \sigma}{9}} - \nn\\
 & &  {\frac{{\sqrt{1 - 4\,z}}\,\left( 1 + 2\,z \right) \,
      \left( 2\,\log x_1 - \log z \right) }{9}}\,,\nn\\
P^{cc}_{2,22}&=&{\frac{16\,{\sqrt{1 - 4\,z}}\,\left( 1 + 8\,z + 12\,{z^2} \right) }{27}} +
  {\frac{8\,\log \sigma}{9}} - {\frac{32\,{z^2}\,\log \sigma}{3}} -
  {\frac{128\,{z^3}\,\log \sigma}{9}} + \nn\\
 & & {\frac{8\,{\sqrt{1 - 4\,z}}\,\left( 1 + 2\,z \right) \,
      \left( 2\,\log x_1 - \log z \right) }{9}}\,,
\eea

\end{footnotesize}

\begin{footnotesize}
\be
\begin{array}{lll}
  P^c_{1,13}=3\,{\sqrt{1 - 4\,z}}\,\left( 1 - z \right) \,, \qquad
& P^c_{1,23}= {\sqrt{1 - 4\,z}}\,\left( 1 - z \right) \,,\qquad
& P^c_{1,14}= {\sqrt{1 - 4\,z}}\,\left( 1 - z \right) \,,\\
  P^c_{2,13}=3\,{\sqrt{1 - 4\,z}}\,\left( 1 + 2\,z \right)\,, \qquad
& P^c_{2,23}={\sqrt{1 - 4\,z}}\,\left( 1 + 2\,z \right)\,, \qquad
& P^c_{2,14}={\sqrt{1 - 4\,z}}\,\left( 1 + 2\,z \right) \,.
\end{array}
\ee

\be
\begin{array}{lll}
  P^c_{1,24}={\frac{1}{2}}\,{\left( 1 - 4\,z \right) }^{\frac{3}{2}}\,, \qquad
& P^c_{1,15}=9\,z\,{\sqrt{1 - 4\,z}}\,,\qquad
& P^c_{1,25}=3\,z\,{\sqrt{1 - 4\,z}}\,,\\
  P^c_{2,24}=- {\sqrt{1 - 4\,z}}\,\left( 1 + 2\,z \right)  \,, \qquad
& P^c_{2,15}=0\,, \qquad
& P^c_{2,25}=0\,.
\end{array}
\ee

\be
\begin{array}{lll}
  P^c_{1,16}=3\,z\,{\sqrt{1 - 4\,z}}\,, \qquad
& P^c_{1,26}=3\,z\,{\sqrt{1 - 4\,z}}\,,\qquad
& P^c_{1,28}=-\,{\frac{1}{6}}\,{\sqrt{1 - 4\,z}}\,\left( 1 + 2\,z \right)\,,\\
  P^c_{2,16}=0\,, \qquad
& P^c_{2,26}=0\,, \qquad
& P^c_{2,28}={\frac{4}{3}}\,{\sqrt{1 - 4\,z}}\,\left( 1 + 2\,z \right)\,.
\end{array}
\ee
\end{footnotesize}
The functions $P^u_{k,ij}$ can be obtained from $P^c_{k,ij}$ by taking the limit
$z\to 0$.

\section*{Appendix B: Evanescent Operators and Renormalization Schemes}
\label{sec:Evan}

In this appendix, we define the $\msb$ scheme chosen to renormalize the QCD,
$\DB=2$ operators whose Wilson coefficients are given in appendix A. This scheme is
introduced in ref.~\cite{Beneke:1998sy} by giving three prescriptions
(eqs.~(13)-(15) of that paper) which define implicitly the relevant evanescent 
operators. In order to make it more explicit, we write out in this appendix
the complete basis of operators and evanescent operators defining this scheme.
Although this renormalization scheme is the only one needed to define our
results, we think it may be useful, for future applications, to present also the
matching matrix which relates the QCD operators in the $\msb$ of
\cite{Beneke:1998sy} to a corresponding set of HQET operators. This matrix has
been computed and used in the intermediate steps of our calculation. To this
purpose, we consider a $\msb$ scheme for the HQET operators which is a simple
generalization of the one defined in ref.~\cite{Gimenez:1998mw}. In addition,
since a second $\msb$ renormalization scheme has been introduced in ref.~\cite{db2nlo1}
for the $\DB=2$ operators in QCD, we also discuss this scheme and present the
corresponding NLO matching matrix relating the QCD operators to the HQET ones.

To start with, we introduce the following set of $\Delta B=2$ four-fermion
operators,
\renewcommand{\arraystretch}{2.0} \be
\begin{array}{ll}
{ O}^q_1 = (\bar b_i q_i)_{V-A} (\bar b_j q_j)_{V-A}\,,&
{ O}^q_2 = (\bar b_i q_i)_{S-P} (\bar b_j q_j)_{S-P}\,,\cr
{ O}^q_3 = (\bar b_i\, \sigma^{\mu\nu}_L\, q_i) (\bar
b_j\,\sigma_{\mu\nu\,L}\, q_j)\,,&
{ O}^q_4 = (\bar b_i\, q_j)_{S-P} (\bar b_j\, q_i)_{S-P}\,,\cr
\end{array}
\label{eq:effop2}
\ee
where, for any string $\Gamma$ of Dirac matrices, we define $\Gamma_L=
\Gamma (1-\gamma_5)$.

Concerning the evanescent operators, it is worth to recall that they are
renormalized, in any given renormalization scheme, in such a way that their
matrix elements vanish on IR finite, physical external states. Consequently, the
anomalous dimension matrix elements mixing the physical and the evanescent
operators vanish to all orders~\cite{grinstein}. Furthermore, while the
evanescent operators usually only enter the definition of the renormalization
scheme, in some cases they can also contribute beyond the LO to the matching of
the physical operators. In particular, when the matching is performed in the
presence of IR divergences, as in our calculation, one should properly take into
account the contribution of the matrix elements of the evanescent operators to
the matching conditions of the physical operators~\cite{Ciuchini:2001vx,Misiak:1999yg}.

We now proceed by defining in details the several renormalization schemes
discussed above.

\subsection*{\boldmath$\msb$ scheme for QCD operators of
ref.~\cite{Beneke:1998sy}}

This scheme concerns $\DB=2$ operators in QCD. The four-dimensional basis
involves the operators $O^q_1$, $O^q_2$ and $O^q_4$ in
eq.~(\ref{eq:effop2}), whereas the evanescent operators are
\bea\renewcommand{\arraystretch}{2.8}
S^q_1 &=&  (\bar b_i\, q_j)_{V-A} (\bar b_j\, q_i)_{V-A}-  { O}^q_1 \,,\nn\\
S^q_2 &=&  (\bar b_i\, \g^\mu\g^\nu\g^\rho_L\, q_j) (\bar b_j\, \g_\mu\g_\nu
\g_{\rho\,L}\, q_i)-\left(16-4\,\ep\right)\left(S^q_1+{ O}^q_1 \right)\,,
\nn\\
S^q_3 &=&  (\bar b_i\, \sigma^{\mu\nu}_L\, q_j) (\bar b_j\,\sigma_{\mu\nu\,L}\,
q_i) -(4-2\,\ep)\,{ O}^q_4-( 8 -8\,\ep)\,{ O}^q_2 \,,\\
S^q_4 &=&  (\bar b_i\, \g^\mu\g^\nu\g^\rho_L\, q_i) (\bar b_j\, \g_\mu\g_\nu
\g_{\rho\,L}\, q_j)- (16-4\ep)\, { O}^q_1 \,,\nn\\
S^q_5 &=&  (\bar b_i\, \sigma^{\mu\nu}_L\, q_i) (\bar b_j\,\sigma_{\mu\nu\,L}
\, q_j) -(4-2\,\ep)\,{ O}^q_2-( 8 -8\,\ep)\,{ O}^q_4 \,.\nn
\label{eq:r0}
\eea

\subsection*{\boldmath$\msb$ scheme for QCD operators of ref.~\cite{db2nlo1}}

This scheme is defined by choosing $O^q_1$,
$O^q_2$ and $O^q_3$ in eq.~(\ref{eq:effop2}) as operators of the
physical basis and the following set of evanescent operators
\bea\renewcommand{\arraystretch}{2.8}
M^q_1 &=&  (\bar b_i\, q_j)_{V-A} (\bar b_j\, q_i)_{V-A}-  { O}^q_1 \,,\nn\\
M^q_2 &=&  (\bar b_i\, \g^\mu\g^\nu\g^\rho_L\, q_j) (\bar b_j\, \g_\mu\g_\nu
\g_{\rho\,L}\, q_i)-\left(16-4\,\ep\right)\left(M^q_1+{ O}^q_1 \right)\,,
\nn\\
M^q_3 &=&  (\bar b_i\, \sigma^{\mu\nu}_L q_j) (\bar b_j\,\sigma_{\mu\nu\,L}q_i)
-  6\,{ O}^q_2  -  { O}^q_3/2\,,\nn\\
M^q_4 &=&  (\bar b_i\, q_j)_{S-P} (\bar b_j\, q_i)_{S-P}+  { O}^q_2/2 -
{ O}^q_3/8 \,,\\
M^q_5 &=&  (\bar b_i\, \g^\mu\g^\nu\g^\rho\g^\sigma_L\, q_j) (\bar b_j\, \g_\mu
\g_\nu\g_\rho\g_{\sigma\,L}\, q_i)- 16 \, M^q_3 - 64\, M^q_4
- 64\, { O}^q_2 -16\,\left(1 - \ep\right)\, { O}^q_3\,,\nn\\
M^q_6 &=&  (\bar b_i\, \g^\mu\g^\nu\g^\rho_L\, q_i) (\bar b_j\, \g_\mu\g_\nu
\g_{\rho\,L}\, q_j)-\left(16-4\,\ep\right)\,{ O}^q_1 \,,\nn\\
M^q_7 &=&  (\bar b_i\, \g^\mu\g^\nu\g^\rho\g^\sigma_L\, q_i) (\bar b_j\,
\g_\mu\g_\nu\g_\rho\g_{\sigma\,L}\, q_j) -\left(64-96\,\ep\right){ O}^q_2 -
\left(16-8\,\ep\right){ O}^q_3\,.\nn
\eea

\subsection*{\boldmath$\msb$ scheme for HQET operators}

In the $\DB=2$ HQET Hamiltonian, we choose the operators $O^q_1$ and
$O^q_2$ in eq.~(\ref{eq:effop2}) as operators of the physical basis and
the following set of evanescent operators,
\bea\renewcommand{\arraystretch}{2.8}
H^q_1 &=&  (\bar b_i\, q_j)_{V-A} (\bar b_j\, q_i)_{V-A}-  { O}^q_1 \nn\\
H^q_2 &=&  (\bar b_i\, \g^\mu\g^\nu\g^\rho_L\, q_j) (\bar b_j\, \g_\mu\g_\nu
\g_{\rho\,L}\, q_i)-\left(16-4\,\ep\right)\left(H^q_1+{ O}^q_1 \right)\nn\\
H^q_3 &=&  (\bar b_i\, \sigma^{\mu\nu}_L q_j) (\bar b_j\,\sigma_{\mu\nu\,L} q_i)
+ \left(4-2\,\ep\,\tau\right)\left(H^q_1+{ O}^q_1\right)+\left(4-2\,\ep\,
\sigma\right)\left( H^q_4 -{ O}^q_1/2 -{ O}^q_2\right)\nn\\
H^q_4 &=& { O}^q_4  + { O}^q_1/2 +{ O}^q_2   \\
H^q_5 &=&  (\bar b_i\, \g^\mu\g^\nu\g^\rho_L\, q_i) (\bar b_j\, \g_\mu\g_\nu
\g_{\rho\,L}\, q_j)-\left(16-4\,\ep\right)\, { O}^q_1 \nn\\
H^q_6 &=& { O}^q_3 + \left(4-2\,\ep\,\tau\right){
O}^q_1+\left(4-2\,\ep\,\sigma\right) { O}^q_2\nn
\label{eq:Heva}
\eea
where $O^q_3$ and $O^q_4$ are also defined in
eq.~(\ref{eq:effop2}). The parameters $\sigma$ and $\tau$ are not fixed and
enter the definition of the renormalization scheme. For the specific choice
$\sigma=\tau$ this scheme reduces to the one considered in
ref.~\cite{Gimenez:1998mw}. In four dimensions the operators $H^q_1-H^q_6$
vanish because of the Fierz identities, the Chisholm identity
\be
\g_\mu\g_\nu\g_\rho=g_{\mu\nu}\,\g_\rho+g_{\nu\rho}\g_\mu-
g_{\mu\rho}\g_\nu
+i\,\ep_{\alpha\mu\nu\rho}\,\g^\alpha\g_5
\ee
and the equation $\g^0 b=b$, verified by the $b$-field operator in the static
limit. Notice that, since $H^q_4$ and $H^q_6$ are evanescent, both
$O^q_3$ and $O^q_4$ in the HQET are reducible in terms of $O^q_1$ and
$O^q_2$ and need not to be included in the basis.

\subsection*{Matching between QCD and HQET operators}

We now discuss the matching, at the NLO, between QCD and HQET operators. The 
matching condition can be written as
\be
\label{eq:matchQHET}
O_k(m_b)_{\rm QCD} =
C_{kl}\,O_l(m_b)_{\rm HQET}\, +{\cal O}(1/m_b),
\ee
where a common renormalization scale $\mu=m_b$ has been chosen. Let us stress
that the equation above is only valid once the operators are sandwiched between
external on-shell states.

For the QCD $\msb$ scheme of ref.~\cite{Beneke:1998sy}, the matrix $\widehat C$
has been given at ${\cal O}(\as)$ in ref.~\cite{Becirevic:2001xt}, by
considering in the HQET the renormalization scheme defined above with the
particular choice $\sigma=\tau=0$. We have repeated the calculation for generic 
$\sigma$ and $\tau$ and obtained
\renewcommand{\arraystretch}{1.0}
\be\renewcommand{\arraystretch}{1.0}
\left(\begin{array}{c}
{ O}_1(m_b)\cr
{ O}_2(m_b)\cr
{ O}_4(m_b)\cr
\end{array}\right)^{\rm{ref.}~\cite{Beneke:1998sy}}_{\rm QCD}
=\left[
\left(\begin{array}{ccc}
   1 & 0 \cr
   0 & 1 \cr
  -1/2 & -1 \cr
\end{array}\right)+
\dfrac{\as(m_b)}{4\pi}\,D_1\right]
\,\left(\begin{array}{c}
{ O}_1(m_b)\cr
{ O}_2(m_b)\cr
\end{array}\right)_{\rm HQET},
\label{eq:Mben}
\ee
\renewcommand{\arraystretch}{2.0}
\be
D_1=
\left(\begin{array}{ccc}
   -14 & -8  \cr
   \dfrac{39-3\sigma+4\tau}{24} & \dfrac{25-\sigma}{3}  \cr
   \dfrac{121+3\sigma-4\tau}{24} & \dfrac{11+\sigma}{3}  \cr
\end{array}\right).
\label{eq:Mben2}
\ee
For $\sigma=\tau=0$ this confirms the previous results.

Notice that in QCD the operator $O^q_4$ is in general independent from
the operators $O^q_1$ and $O^q_2$. However, its matrix elements
between external on-shell states can be expressed in terms of the matrix
elements of $O^q_1$ and $O^q_2$. This relation has been given in
ref.~\cite{Beneke:1998sy} in terms of the operator
\be
R^q_0 ={ O}^q_4 +{ O}^q_1/2+{ O}^q_2 \, .
\ee
The on-shell matrix elements of $R^q_0$ are of short-distance origin and their
values can be easily obtained by using the matching coefficients between QCD and
HQET operators given in eq.~(\ref{eq:Mben}).

When the QCD operators are renormalized in the $\msb$ of ref.~\cite{db2nlo1} 
the matrix $\widehat C$ is given by
\be\renewcommand{\arraystretch}{1.5}
\left(\begin{array}{c}
{ O}_1(m_b)\cr
{ O}_2(m_b)\cr
{ O}_3(m_b)\cr
\end{array}\right)^{\rm{ref.}~\cite{db2nlo1}}_{\rm QCD}
=\left[\left(\begin{array}{ccc}
   1 & 0 \cr 
   0 & 1 \cr
   4 & 4 \cr
\end{array}\right)+\dfrac{\as(m_b)}{4\pi}\, D_1\right]
\left(\begin{array}{c}
{ O}_1(m_b)\cr
{ O}_2(m_b)\cr
\end{array}\right)_{\rm HQET},
\label{eq:Mmis}
\ee
\renewcommand{\arraystretch}{2.0}
\be
D_1=
\left(\begin{array}{ccc}
   -14 & -8  \cr
   \dfrac{35-3\sigma+4\tau}{24} & \dfrac{13-\sigma}{3}  \cr
   \dfrac{237+3\sigma-4\tau}{6} & 4\dfrac{27+\sigma}{3}  \cr
\end{array}\right).
\label{eq:Mmis2}
\ee
By using the above results one also finds that in this scheme the matching of
the operator ${ O}_4$ onto the HQET operators is given by
\be\renewcommand{\arraystretch}{0.5}
\left(\begin{array}{c}
{ O}_4(m_b)\cr
\end{array}\right)^{\rm{ref.}~\cite{db2nlo1}}_{\rm QCD}
=\left[
\left(\begin{array}{ccc}
   -1/2 & -1 \cr
\end{array}\right)+
\dfrac{\as(m_b)}{4\pi}\,
\left(\begin{array}{ccc}
    \dfrac{101+3\sigma-4\tau}{24} & \dfrac{7+\sigma}{3}  \cr
\end{array}\right) \right]
\,\left(\begin{array}{c}
{ O}_1(m_b)\cr
{ O}_2(m_b)\cr
\end{array}\right)_{\rm HQET}.
\label{eq:M4mis}
\ee
\renewcommand{\arraystretch}{2.0}



\begin{thebibliography}{99}

\bibitem{Battaglia:2003in}
M.~Battaglia {\it et al.},
arXiv:hep-ph/0304132.\\
See also The Heavy Flavor Averaging Group (HFAG),
http://www.slac.stanford.edu/ xorg/hfag/

\bibitem{Aubert:2003se}
B.~Aubert {\it et al.}  [BABAR Collaboration],
arXiv:hep-ex/0303043.

\bibitem{Ciuchini:2003rk}
M.~Ciuchini, E.~Franco, F.~Parodi, V.~Lubicz, L.~Silvestrini and A.~Stocchi,
arXiv:hep-ph/0307195.

\bibitem{Khoze:1987fa}
V.A.~Khoze, M.A.~Shifman, N.G.~Uraltsev and M.B.~Voloshin,
Sov.\ J.\ Nucl.\ Phys.\  {\bf 46} (1987) 112
[Yad.\ Fiz.\  {\bf 46} (1987) 181].

\bibitem{ope}
J.~Chay, H.~Georgi and B.~Grinstein,
Phys.\ Lett.\ B {\bf 247} (1990) 399.

\bibitem{Ciuchini:2001vx}
M.~Ciuchini, E.~Franco, V.~Lubicz and F.~Mescia,
Nucl.\ Phys.\ B {\bf 625}, 211 (2002)
[hep-ph/0110375].

\bibitem{Beneke:2002rj}
M.~Beneke, G.~Buchalla, C.~Greub, A.~Lenz and U.~Nierste,
Nucl.\ Phys.\ B {\bf 639}, 389 (2002)
[hep-ph/0202106].

\bibitem{Franco:2002fc}
E.~Franco, V.~Lubicz, F.~Mescia and C.~Tarantino,
Nucl.\ Phys.\ B {\bf 633}, 212 (2002)
[hep-ph/0203089].

\bibitem{Gabbiani:2003pq}
F.~Gabbiani, A.~I.~Onishchenko and A.~A.~Petrov,
arXiv:hep-ph/0303235.

\bibitem{Beneke:1996gn}
M.~Beneke, G.~Buchalla and I.~Dunietz,
Phys.\ Rev.\ D {\bf 54}, 4419 (1996)
[hep-ph/9605259].

\bibitem{Beneke:1998sy}
M.~Beneke, G.~Buchalla, C.~Greub, A.~Lenz and U.~Nierste,
Phys.\ Lett.\ B {\bf 459}, 631 (1999)
[hep-ph/9808385].

\bibitem{Dighe:2001gc}
A.~S.~Dighe, T.~Hurth, C.~S.~Kim and T.~Yoshikawa,
Nucl.\ Phys.\ B {\bf 624}, 377 (2002)
[hep-ph/0109088].

\bibitem{Becirevic:2001xt}
D.~Becirevic, V.~Gimenez, G.~Martinelli, M.~Papinutto and J.~Reyes,
JHEP {\bf 0204}, 025 (2002)
[hep-lat/0110091].

\bibitem{hagelin}
J.~S.~Hagelin,
Nucl.\ Phys.\ B {\bf 193}, 123 (1981).

\bibitem{Buras:1990fn}
A.~J.~Buras, M.~Jamin and P.~H.~Weisz,
Nucl.\ Phys.\ B {\bf 347}, 491 (1990).

\bibitem{nlodb1a}
A.J.~Buras, M.~Jamin, M.E.~Lautenbacher and P.H.~Weisz,
Nucl.\ Phys.\ B {\bf 400} (1993) 37
[hep-ph/9211304].

\bibitem{nlodb1b}
A.J.~Buras, M.~Jamin and M.E.~Lautenbacher,
Nucl.\ Phys.\ B {\bf 400} (1993) 75
[hep-ph/9211321].

\bibitem{nlodb1c}
M.~Ciuchini, E.~Franco, G.~Martinelli and L.~Reina,
Nucl.\ Phys.\ B {\bf 415} (1994) 403
[hep-ph/9304257].

\bibitem{tarcer}
R.~Mertig and R.~Scharf,
Comput.\ Phys.\ Commun.\  {\bf 111} (1998) 265
[hep-ph/9801383].

\bibitem{Misiak:1999yg}
M.~Misiak and J.~Urban,
Phys.\ Lett.\ B {\bf 451} (1999) 161
[hep-ph/9901278].

\bibitem{db2nlo1}
A.~J.~Buras, M.~Misiak and J.~Urban,
Nucl.\ Phys.\ B {\bf 586}, 397 (2000)
[hep-ph/0005183].

\bibitem{Gimenez:1998mw}
V.~Gimenez and J.~Reyes,
Nucl.\ Phys.\ B {\bf 545} (1999) 576
[arXiv:hep-lat/9806023].

\bibitem{Gimenez:2000jj}
V.~Gimenez and J.~Reyes,
Nucl.\ Phys.\ Proc.\ Suppl.\  {\bf 94} (2001) 350
[hep-lat/0010048].

\bibitem{Hashimoto:2000eh}
S.~Hashimoto, K.~I.~Ishikawa, T.~Onogi, M.~Sakamoto, N.~Tsutsui and N.~Yamada,
Phys.\ Rev.\ D {\bf 62} (2000) 114502
[hep-lat/0004022].

\bibitem{Aoki:2002bh}
S.~Aoki {\it et al.}  [JLQCD Collaboration],
Phys.\ Rev.\ D {\bf 67} (2003) 014506
[hep-lat/0208038].

\bibitem{Becirevic:2000sj}
D.~Becirevic, D.~Meloni, A.~Retico, V.~Gimenez, V.~Lubicz and G.~Martinelli,
Eur.\ Phys.\ J.\ C {\bf 18} (2000) 157
[hep-ph/0006135].

\bibitem{Lellouch:2000tw}
L.~Lellouch and C.~J.~Lin  [UKQCD Collaboration],
Phys.\ Rev.\ D {\bf 64} (2001) 094501
[hep-ph/0011086].

\bibitem{Yamada:2001xp}
N.~Yamada {\it et al.}  [JLQCD Collaboration],
Nucl.\ Phys.\ Proc.\ Suppl.\  {\bf 106} (2002) 397
[hep-lat/0110087].

\bibitem{Aoki:2003xb}
S.~Aoki {\it et al.}  [JLQCD Collaboration],
arXiv:hep-ph/0307039.

\bibitem{Korner:2003zk}
J.~G.~Korner, A.~I.~Onishchenko, A.~A.~Petrov and A.~A.~Pivovarov,
arXiv:hep-ph/0306032.

\bibitem{Hagiwara:2002hf}
K.~Hagiwara, S.~Narison and D.~Nomura,
Phys.\ Lett.\ B {\bf 540}, 233 (2002)
[hep-ph/0205092].

\bibitem{Narison:1994zt}
S.~Narison and A.~A.~Pivovarov,
Phys.\ Lett.\ B {\bf 327}, 341 (1994)
[hep-ph/9403225].

\bibitem{grinstein}
M.~J.~Dugan and B.~Grinstein,
Phys.\ Lett.\ B {\bf 256} (1991) 239.

\bibitem{Beneke:2003az}
M.~Beneke, G.~Buchalla, A.~Lenz and U.~Nierste,
arXiv:hep-ph/0307344.

\end{thebibliography}
\end{document}